\documentclass{spi-b3068}
\usepackage{multirow}

\makeindex

\begin{document}

\chapter[Hydrodynamic methods and sub-resolution models for cosmological simulations]{Hydrodynamic methods and sub-resolution models \break for cosmological simulations}

\author[M. Valentini \& K. Dolag]{Milena Valentini$^{\ast, \dag, \triangle}$ and Klaus Dolag$^{\S,\ddag,\P}$}

\address{$^{\ast}\!$Dipartimento di Fisica, Università degli Studi di Trieste\\
Via Alfonso Valerio 2, 34127 Trieste, Italy\\
$^{\dag}\!$INAF - Astronomical Observatory of Trieste\\ 
via Tiepolo 11, I-34131 Trieste, Italy\\
$^{\S}\!$Universit\"ats-Sternwarte M\"unchen, LMU\\ Scheinerstr.\ 1 D-81679 M\"unchen, Germany\\
$^{\ddag}\!$MPI for Astrophysics\\ 
Karl-Schwarzschild Strasse 1 D-85748 Garching, Germany\\
$^{\triangle}\!$milena.valentini@units.it\\
$^{\P}\!$dolag@usm.uni-muenchen.de}

\begin{abstract}
Cosmological simulations are powerful tools in the context of structure formation. They allow us to explore the hierarchical assembly of dark matter (DM) halos and their clustering, to validate or reject possible scenarios of structure formation, and to investigate the physical properties of evolving galaxies across cosmic time. Cosmological hydrodynamical simulations are especially key to study how the complex interstellar medium (ISM) of forming galaxies responds to the most energetic processes during galaxy evolution, such as stellar feedback ensuing supernova (SN) explosions and feedback from AGN (active galactic nuclei). 
Given the huge dynamical range of physical scales spanned by the astrophysical processes involved in cosmic structure formation and evolution, cosmological simulations resort to sub-resolution models to capture processes occurring below their resolution limit. The impact of different sub-grid prescriptions accounting for the same process is striking, though often overlooked.
Some among the main aforementioned processes include: hot gas cooling, star formation and stellar feedback, stellar evolution and chemical enrichment, black hole (BH) growth and their ensuing feedback.
Producing simulations of cosmic structure formation and galaxy evolution in large computational volumes is key to shed new light on what drives the formation of the first structures in the Universe, and their subsequent evolution. Not only are predictions from simulations crucial to compare with data from ongoing and upcoming observational instruments, but they can also effectively guide future observational campaigns. 
Besides, since we have entered the era of high-performance computing, it is of paramount importance to have numerical codes which are not only as complete as possible as for the inclusion of physical processes implemented, but also very efficient from the computational point of view and able to smoothly scale on state-of-the-art exascale infrastructures. 
In this chapter, we review the main hydrodynamic methods used in cosmological simulations and the most common techniques adopted to include the fundamental astrophysical processes which drive galaxy formation and evolution.
\end{abstract}

\section{Cosmological hydrodynamical simulations}\label{ch3_intro}

Being gravity the force that drives structure formation, the building block of the majority of cosmological simulations is an $N$-body code (see e.g. \cite{Yoshikawa2013} for a discussion on alternative approaches). The goal of an N-body code is to investigate the non-linear dynamical evolution of a self-gravitating, collisionless system (see \cite{Dolag2008, Springel2016}, for reviews). 

A collisionless fluid is a system whose constituent elements move under the effect of the collective gravity field generated by the total mass distribution: in a collisionless system, gravitational collisions and interactions between fluid elements are negligible and do not influence the motion and general properties of the constituent elements themselves. 
In order to establish quantitatively whether a system can be considered collisionless or not, the relaxation timescale has to be contrasted to the timescale over which the system evolves. 
A system featuring $N$ elements moving with typical velocity $v$ across its size $L$ 
is characterised by the crossing time $t_{\rm cross}  = L/v$ \cite{BandT}.
Possible collisions with nearby particles result in a quadratic velocity perturbation $(\Delta v)^2$ of each particle's velocity, since each encounter contributes with a randomly oriented $\Delta v$.
The relaxation time is the timescale over which the quadratic velocity perturbation amounts to roughly the squared velocity itself $t_{\rm relax}  = v^2/(\Delta v)^2 \times t_{\rm cross}$. 
Here, $v^2/(\Delta v)^2$ quantifies the number of crossings required to have a velocity change of the same order as $v^2$ and depends on the number $N$ of particles \cite{BandT}. 
Therefore, the relaxation time can be cast as $t_{\rm relax}  = N/(8\, {\text{ln}} N) \times t_{\rm cross}$.  
A system can be deemed as collisionless if its relaxation time exceeds by far the age of the Universe $\sim 1/H_{\rm 0}$ ($H_{\rm 0}$ being the present-day value of the Hubble constant). After the timescale $t_{\rm relax}$, deflections in a particle motion by interactions with other particles are expected, and the system cannot be considered collisionless anymore. 
DM and stars in galaxies can be represented as a collisionless, non-relativistic fluid made of particles, while gas is a collisional component.

To a first approximation, the formation of cosmic structures can be studied using $N$-body simulations, which only follow the evolution of collisionless particles under gravity. Such simulations have been performed with high resolution for \hbox{individual} objects, like galaxies and galaxy clusters, as well as for very large-scale \hbox{structures}. The numerical methods and different possible approaches used to tackle the problem of integrating the equations of motion of the $N$ particles in the cumulative gravity field are described in \cref{Ch_nbody}. 

However, with the possible \hbox{exception} of gravitational lensing, observations mainly reflect the state of the ordinary (baryonic) matter. Therefore, their interpretation in the framework of cosmic structure evolution requires that we understand the complex, non-gravitational physical processes which determine the evolution of the cosmic baryons. 
How DM, gas and stars co-evolve --~with baryons falling into the potential well of the underlying DM\index{dark matter} distribution, cooling, and finally condensing to form stars 
(see Video~1, page xiii)~-- within the hierarchical formation scenario contributes to the state and composition of the intergalactic and\index{intergalactic medium (IGM)}\index{IGM, see intergalactic medium} intracluster\index{intracluster medium (ICM)}\index{ICM, see intracluster medium} media (IGM and ICM, respectively), and is responsible for feedback in energy and metals,\index{magnetic field} magnetic fields, and high-energy particles. Depending on their origin, these components will be blown out by jets, winds or ram pressure effects and finally mix with the surrounding IGM/ICM. 
While some of these effects will be naturally followed within hydrodynamic simulations (like ram pressure effects), others have to be included in simulations via effective models (like star formation and ensuing feedback\index{feedback}, and chemical pollution of the ISM, IGM and ICM by SNe).

\begin{figure}
 \centerline{\includegraphics[width=1.0\textwidth]{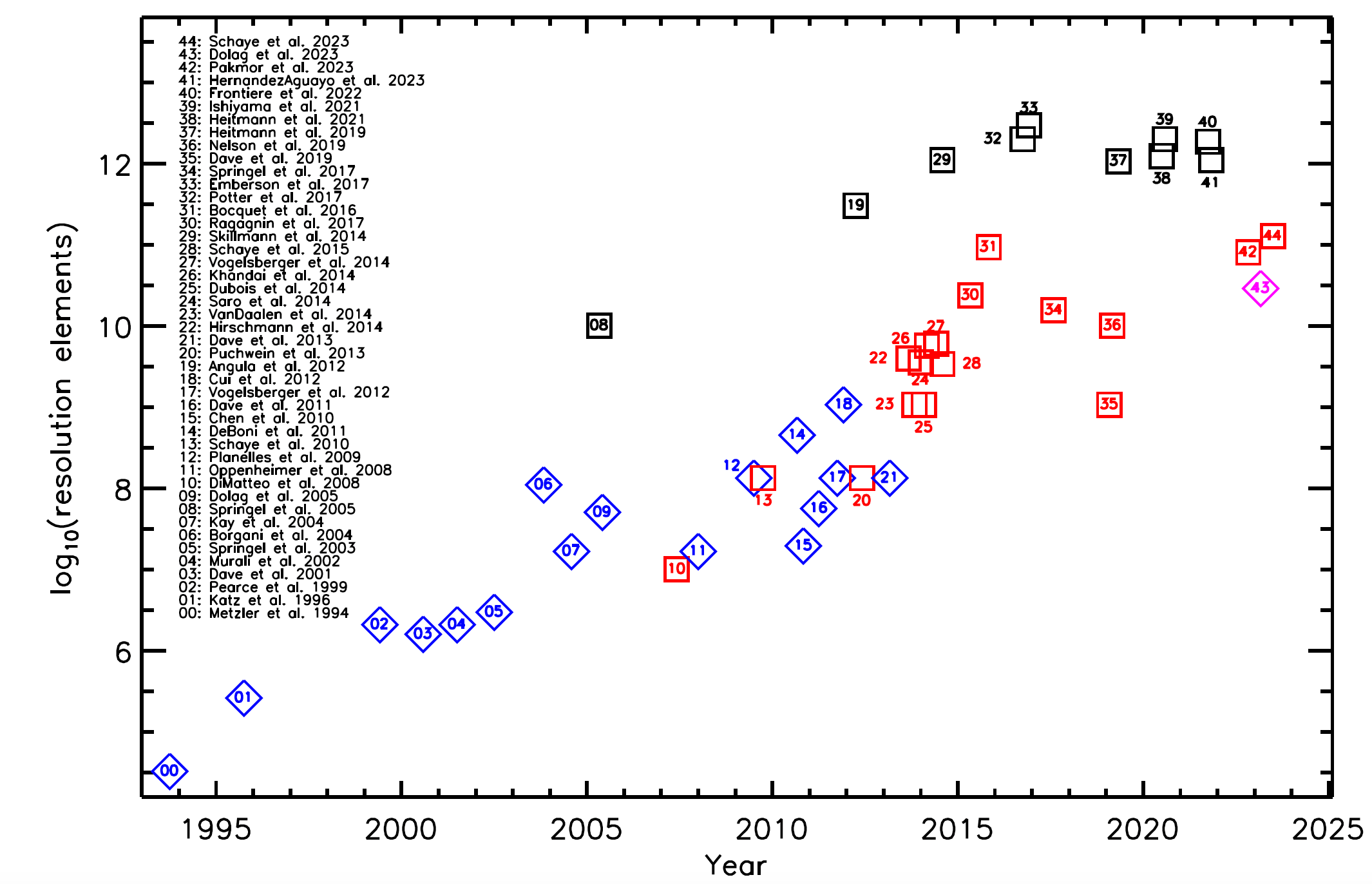}}
 \caption{The evolution of the number of resolution elements in simulations. The black line is a fit to the scaling for pure N-body simulations between 1970 and 2005, taken from \protect{\cite{2005Natur.435..629S}}. The black symbols are recent N-body simulations \protect{\cite{2012MNRAS.426.2046A,2014arXiv1407.2600S,2017ComAC...4....2P,2017RAA....17...85E,2019ApJS..245...16H,2021ApJS..252...19H,2021MNRAS.506.4210I,2022ApJS..259...15F}}.
 The colored symbols pinpoint hydrodynamical, cosmological simulations over the last three decades. Blue data points depict simulations including the effect of cooling and star formation
\protect{\cite{1994ApJ...437..564M, Katz1996, 1999ApJ...521L..99P, dave01, 2002ApJ...571....1M, 2003MNRAS.339..312S, borgani2004, 2004MNRAS.355.1091K, 2005MNRAS.363...29D, 2008MNRAS.387..577O, 2009MNRAS.399..410P, 2011MNRAS.415.2758D, 2011MNRAS.415...11D, 2012MNRAS.425.3024V, 2012MNRAS.423.2279C, 2013MNRAS.434.2645D}}, 
while red data points show simulations which also feature the effect of AGN feedback
\protect{\cite{2008ApJ...676...33D, 2010MNRAS.402.1536S, 2013MNRAS.428.2966P, Hirschmann, Dubois2014, 2014MNRAS.440.2997V, 2014MNRAS.440.2610S, Khandai2015, Vogelsberger2014, Schaye2015, 2016MNRAS.456.2361B, 2016arXiv161206380R, 2019MNRAS.490.3234N, Dave2019, 2023MNRAS.524.2539P, Schaye2023}}. The pink symbol marks a simulation which in addition includes the treatment of magnetic field and spectral cosmic ray electrons and protons \protect{\cite{2023arXiv230210960D}}.
\label{fig:sims}}
\end{figure}

Thanks to the improved computing power and advancements in numerical methods, the number of resolution elements\footnote{To be consistent among different simulation techniques, we define resolution elements here as the number of DM particles only, instead of the total number of particles or cells, which would be typically larger by at least a factor 2.} which can be used in such simulations has increased dramatically over the last decades, as shown in \fref{fig:sims}. 
Note that the largest, hydrodynamical simulations up to date (e.g. {\it Magneticum\footnote{www.magneticum.org} Box0/mr} ~\cite{2016MNRAS.456.2361B} and {\it FLAMINGO\footnote{https://flamingo.strw.leidenuniv.nl/} L2p8\_m9} ~\cite{Schaye2023}) followed a total number of more than $2\times10^{11}$ particles (e.g., DM, gas, stars and BH particles) across the entire evolution of the Universe.

In this chapter, we will review the basic numerical methods which are used to describe the hydrodynamics, and discuss the main approaches adopted to incorporate the fundamental astrophysical processes in the context of cosmological hydrodynamical simulations (see Video~3, page xiii).
Further components like magnetic fields and high-energy particles need additional modelling of their injection processes and evolution. Therefore, they must be self-consistently coupled with hydrodynamics and are described in more detail in \cref{Ch_cluster}.

\enlargethispage{12pt}

\section{Hydrodynamics and numerical methods}

As a result of the high non-linearity of the gravitational clustering in the Universe, numerical simulations are the only method to follow the evolution of the forming structures like galaxies and galaxy clusters in detail. However, a huge dynamic range in space and time has to be captured. For instance, the range of physical scales over which the hierarchical assembly of structures develops spans from sub-kpc scales in galaxies up to several hundreds of megaparsecs, the latter being the largest coherent scale in the Universe. Here, modern cosmological simulation codes based on $N$-body and hydrodynamics techniques are best suited to accurately follow the joint dynamics of DM and gas in their full complexity, during the hierarchical build-up of these structures.

\begin{figure}
\begin{tabular}{m{0.25\textwidth}m{0.25\textwidth}m{0.25\textwidth}m{0.25\textwidth}}
    \multicolumn{2}{c|}{Lagrangian} & \multicolumn{2}{c}{Eulerian} \\
    \includegraphics[width=0.25\textwidth]{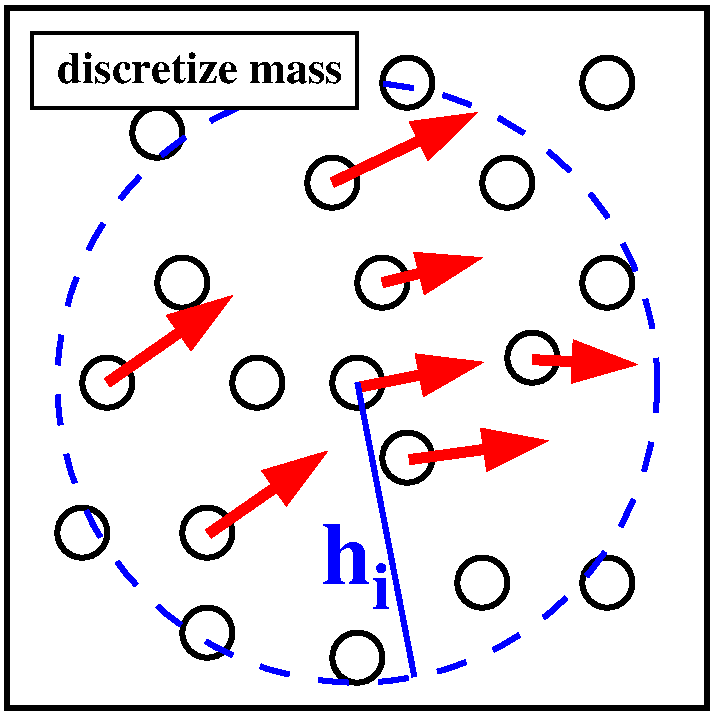} & 
    \multicolumn{2}{c}{$\displaystyle\underbrace{\frac{\mathrm{d}}{\mathrm{d}t}}_{\substack{\text{Lagrangian}\\ \text{derivative}}} = \overbrace{\underbrace{\frac{\partial}{\partial t}}_{\substack{\text{Eulerian}\\ \text{derivative}}}}^{\substack{\text{change at}\\ \text{fluid location}}} + 
    \overbrace{
    \underbrace{\vec{v}\cdot\vec{\nabla}}_{\substack{\text{convective}\\ \text{derivative}}}
    }^{\substack{\text{fluid element}\\ \text{moves}}}
    $} & 
    \includegraphics[width=0.25\textwidth]{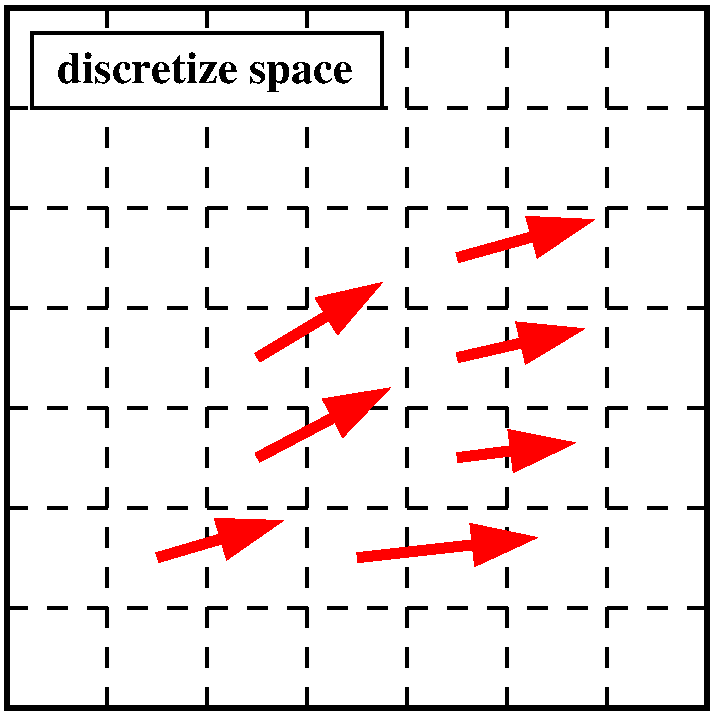} \\
    \hline
    \multicolumn{4}{c}{Continuity equation}\\
    \multicolumn{2}{c|}{$\displaystyle\underbrace{\frac{\mathrm{d}\rho}{\mathrm{d} t}}_{\substack{\text{change of}\\ \text{density}}} + ~~\rho\!\!\!\underbrace{{\vec \nabla}\cdot{\vec v}}_{\substack{\text{divergence}\\ \text{of velocity}}} = 0$} & 
    \multicolumn{2}{c}{$\displaystyle\frac{\partial\rho}{\partial t} + \underbrace{{\vec \nabla}\cdot(\rho{\vec v})}_{\substack{\text{divergence}\\ \text{of mass flux}}} = 0$} \\
    \hline
    \multicolumn{4}{c}{Momentum equation}\\
    \multicolumn{2}{c|}{$\displaystyle\underbrace{\rho\frac{\mathrm{d}{\vec v}}{\mathrm{d} t}}_{\substack{\text{change of}\\ \text{momentum}}} = \;- \underbrace{{\vec \nabla} P}_{\substack{\text{pressure}\\ \text{gradient}}} - \underbrace{\rho{\vec \nabla}\Phi}_{\substack{\text{gravitational}\\ \text{force}}}$} & 
    \multicolumn{2}{c}{$\displaystyle\rho\frac{\partial{\vec v}}{\partial t} + \underbrace{\rho{\vec v}({\vec \nabla}\cdot{\vec v})}_{\substack{\text{momentum flux}\\ \text{in and out}}} =\; - {\vec \nabla}P - \rho {\vec \nabla}\Phi$} \\
    \hline
    \multicolumn{4}{c}{First law of thermodynamics}\\
    \multicolumn{2}{c|}{$\displaystyle\underbrace{\frac{\mathrm{d} u}{\mathrm{d} t}}_{\substack{\text{change of}\\ \text{energy per mass}}} \!\!\!\!\!\!=\; - \underbrace{\frac{P}{\rho} {\vec \nabla} \cdot {\vec v}}_{\substack{\text{adiabatic}\\ \text{changes}}} - \underbrace{n^2\frac{\Lambda(u,\rho)}{\rho}}_{\substack{\text{losses by}\\ \text{cooling}}}$} & 
    \multicolumn{2}{c}{$\displaystyle\underbrace{\frac{\partial(\rho u)}{\partial t}}_{\substack{\text{change of}\\ \text{energy per}\\ \text{volume}}} \!\!\!= - \underbrace{{\vec \nabla}\cdot\left[(\rho u + P){\vec v}\right]}_{\substack{\text{divergence}\\ \text{of energy flux}}} \;-\; n^2\Lambda(u,\rho)$} \\
    \hline
    \multicolumn{2}{c|}{Kernel density estimate} & 
    \multicolumn{2}{c}{Reconstruction} \\
    \multicolumn{2}{c|}{\includegraphics[width=0.5\textwidth]{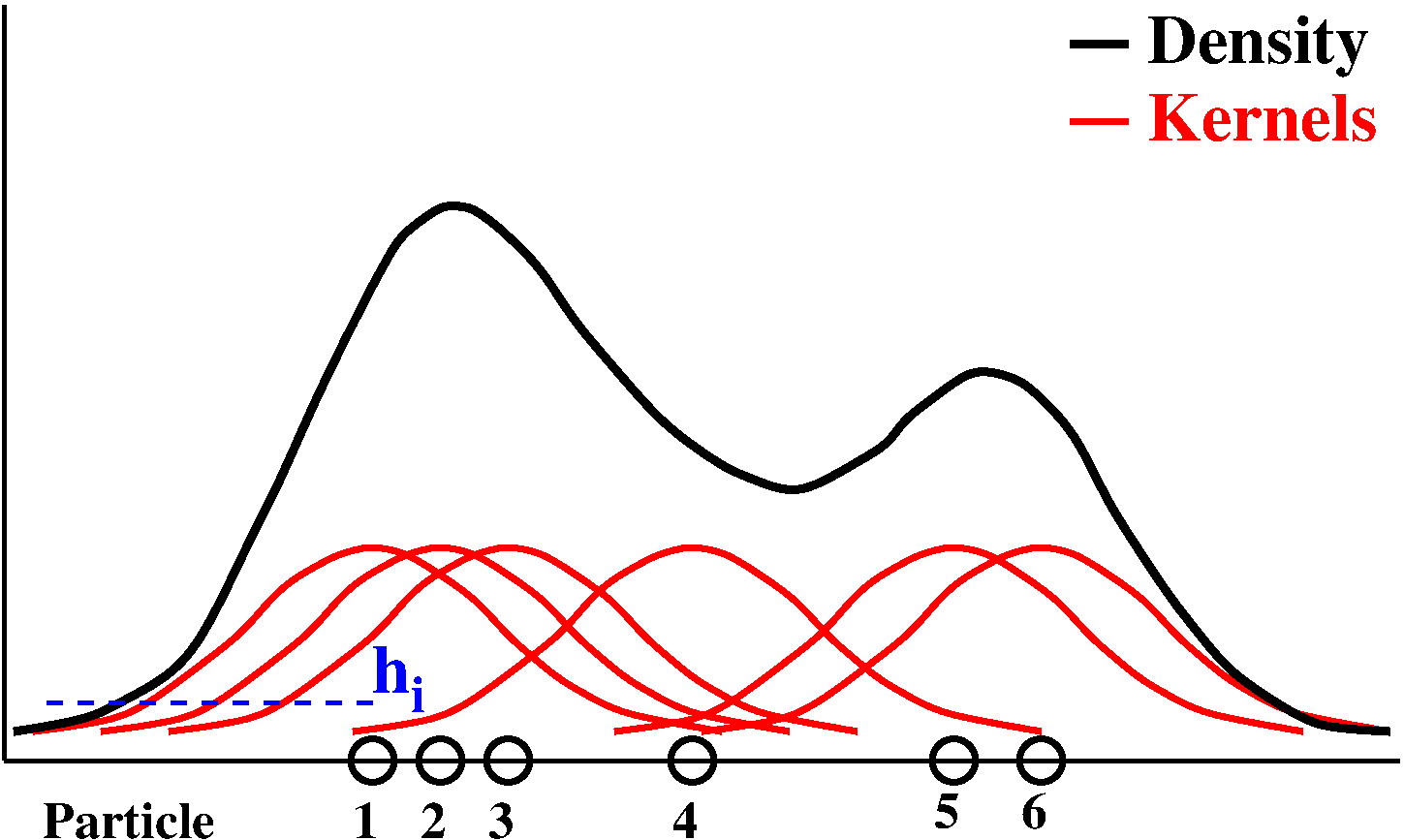}} & 
    \multicolumn{2}{c}{\includegraphics[width=0.5\textwidth]{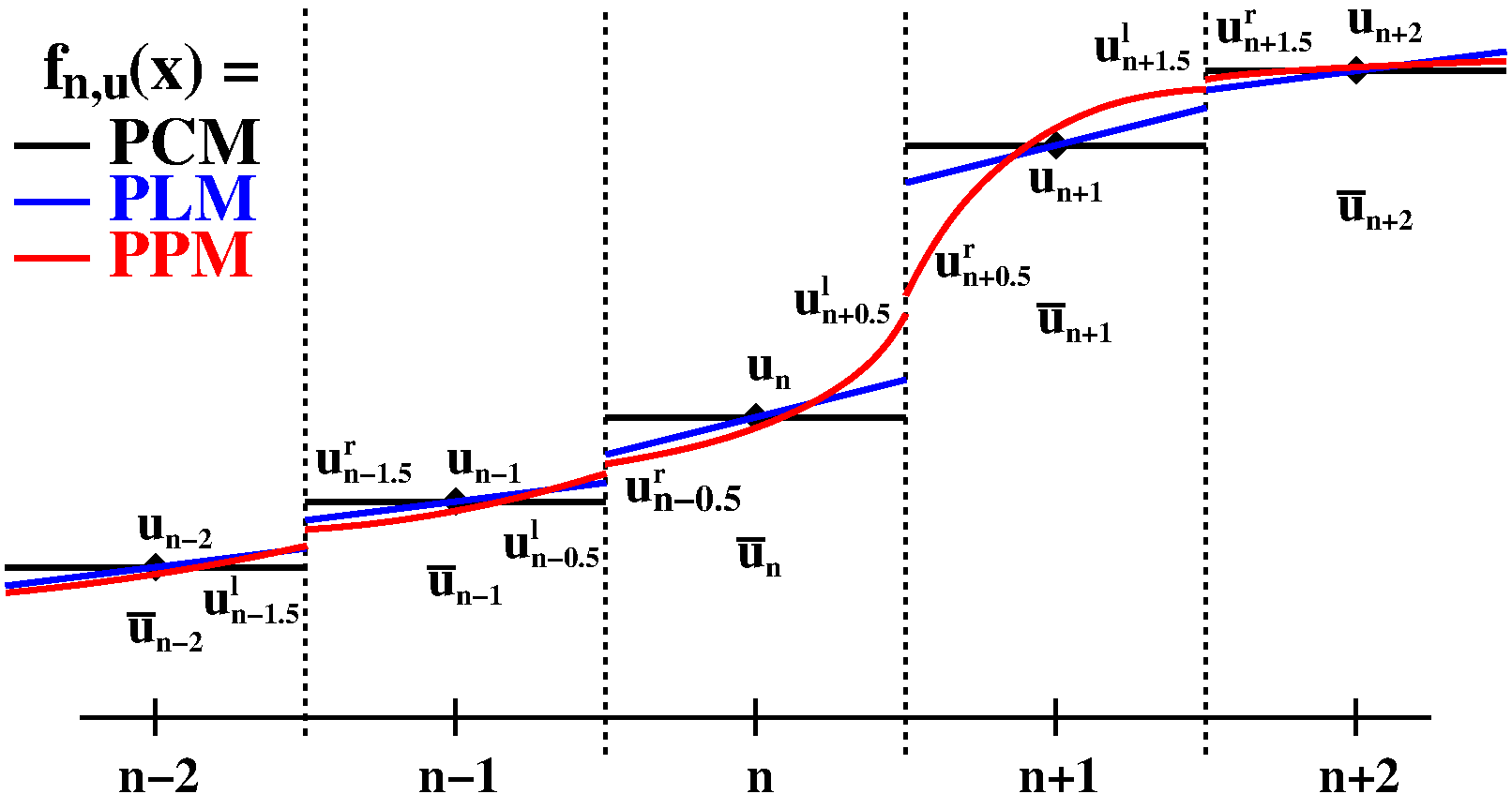}} \\
\end{tabular}
\caption{Comparison of the concept of Lagrangian and Eulerian formulation of hydrodynamics and the resulting set of equations.}
\label{fig:HydroMethods}
\end{figure}

\section{Basic equations and techniques}

A variety of numerical schemes has been developed in the past decades to solve the coupled system of equations describing the baryonic content of the Universe and the collisionless DM\index{dark matter}. 
The majority of the baryons (i.e. gas) can be described as an ideal fluid, whose evolution is ruled by a set of equations, namely, the {\it Euler\index{Euler equations} equations}. 
The hydro solvers which integrate the aforementioned equations fall into two main categories, that are summarized in \fref{fig:HydroMethods}: particle methods, which discretize mass (see \cite{2012JCoPh.231..759P} and references therein), and grid-based methods, which discretize the computational domain (see \cite{2015ARA&A..53..325T} and references therein). 
Recently, additional solvers have been developed: they combine characteristics of both methods (see \cite{2014arXiv1412.5187S} and references therein) and will be discussed in detail in the following sections.

All of them have to be closed by an {\it equation of\index{equation of state} state}, relating the gas pressure $P$ to the internal energy (per unit mass) $u$ and the density $\rho$. Assuming an ideal, monoatomic gas, this will be $P = (\gamma -1)\rho u$ with the polytropic\index{polytropic index} index $\gamma=5/3$. 

When applying these equations to the problem of cosmological structure formation, there are several features emerging in comparison to purely hydrodynamic simulations in a non-expanding background. First, the otherwise often neglected self-gravity, emerging as the ${\vec \nabla}\Phi$ term, has to be accounted for (its contribution can be solved following the methods described in \cref{Ch_nbody}.

Second, radiative losses, quantified through the cooling function $\Lambda(u,\rho)$, where $n^2 \, \Lambda $ ($n^2 \, \Lambda / \rho$) represents the cooling rate per unit volume (mass), with $n$ being the gas number density, play a key role in influencing the evolution of the baryonic component (see also Section~\ref{cooling}). Cooling is especially key to describe the history of star formation of the Universe, as outlined in Section~\ref{sfr}. 

Additionally, the equations have to be adapted to the cosmological background (to take into account the expansion history of the Universe). In almost all cases, the latter step is achieved by switching to the so called comoving coordinates and through a transformation of the time variable. This change does not typically alter the general structure of the equations above, except for additional, cosmological, time dependent pre-factors which appear in the equations (we refer the reader to chapter 5 in \cite{Peebles1994} for a rigorous derivation).

\section{Eulerian $(grid)$ methods}\label{grid}

Grid-based methods solve the {\it Euler equations} (see \fref{fig:HydroMethods}) based on structured or unstructured grids, representing the fluid. 
While {\it primitive variables}\index{primitive variables} encode the thermodynamic properties of the fluid (e.g., $\rho$, ${\vec v}$, or $P$), {\it conservative variables}\index{conservative variables} define the conservation laws (e.g., $\rho$, $\rho{\vec v}$, or $\rho u$). 
Early attempts were made using a central difference scheme, where fluid is only represented by the centered cell values and derivatives are obtained by the finite-difference representation (see, for example, \cite{1992ApJS...78..341C}). These methods use artificial viscosity\index{artificial viscosity} to handle schocks (similar to the smoothed particle hydrodynamics\index{smoothed particle hydrodynamics} method described in Section~\ref{sec:sph}), as they would otherwise break down in regimes where discontinuities appear. Also, by construction, they are only first-order accurate.

Classical approaches use instead reconstruction schemes which, depending on their order, take several neighboring cells (so-called {\it stencils}) into account to reconstruct the field of any hydrodynamical variable ($u_n$), with increasing order of accuracy at the interfaces of the cells ($u_{\pm0.5}^{l/r}$). 
Typical schemes are the {\it piecewise\index{piecewise constant method (PCM)} constant\index{PCM, see piecewise constant method} method} ({\it PCM}), the {\it piecewise linear\index{piecewise linear method (PLM)}\index{PLM, see piecewise linear method} method} ({\it PLM}; e.g., \cite{1985JCoPh..59..264C}), and the {\it piecewise parabolic\index{piecewise parabolic method (PPM)}\index{PPM, see piecewise parabolic method} method} ({\it PPM}; \cite{1984JCoPh..54..174C}), as illustrated in the lower right part of \fref{fig:HydroMethods}. The shape of the reconstruction function is then used to calculate the total integral of a quantity over the grid cell, divided by the volume of each cell (e.g., {\it cell average}, ${\bar u}_n$), rather than pointwise approximations at the grid centers (e.g., {\it central variables, $u_n$}). 

Modern, high-order schemes usually have {\it stencils} based on at least five grid points and implement {\it essentially nonoscillatory} ({\it ENO}; \cite{1987JCoPh..71..231H}) or the so-called {\it weighted essentially\index{weighted  essentially nonoscillatory (WENO)}\index{WENO, see weighted essentially nonoscillatory} nonoscillatory} ({\it WENO}; \cite{1999math.....11089L}) schemes for reconstruction, which maintain high-order accuracy (see e.g., \cite{Shu__1998__CIME__WENO} for a recent review).

\begin{figure}
\begin{tabular}{m{0.5\textwidth}m{0.5\textwidth}}
\multicolumn{1}{c}{Initial state} & \multicolumn{1}{c}{{\it Rankine--Hugoniot} conditions }\\[5pt]
\multicolumn{1}{c}{\multirow{4}{*}{\includegraphics[width=0.5\textwidth]{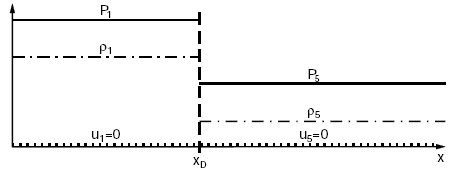}} } &
\multicolumn{1}{c}{$\displaystyle\rho_l v_l = \rho_r v_r$} \\[5pt]
\multicolumn{1}{c}{} &
\multicolumn{1}{c}{$\displaystyle\rho_l v_l^2 + P_l = \rho_r v_r^2 + P_r$} \\[5pt]
\multicolumn{1}{c}{} &
\multicolumn{1}{l}{$\displaystyle v_l (\rho_l(v_l^2/2+u_l) + P_l) =$} \\[4pt]
\multicolumn{1}{c}{} &
\multicolumn{1}{r}{$\displaystyle v_r (\rho_r(v_r^2/2+u_r) + P_r)$ }\\[6pt]
\multicolumn{1}{c}{Final State} & \multicolumn{1}{c}{Solution} \\[5pt]
\multicolumn{1}{c}{\multirow{3}{*}{\includegraphics[width=0.5\textwidth]{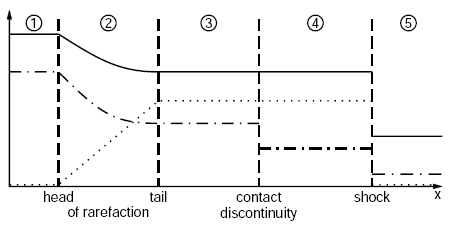}} } &
\multicolumn{1}{l}{$\displaystyle\frac{\rho_1}{\rho_5}\frac{1}{\lambda}\frac{(1-P)^2}{\gamma(1+P)-1+P} =$} \\[12pt]
\multicolumn{1}{c}{} &
\multicolumn{1}{r}{$\displaystyle \;\;\;\;\;\;\;\;\frac{2\gamma}{(\gamma-1)^2}\left[1-\left(\frac{P}{\lambda}\right)^{(\gamma-1)/(2\gamma)}\right]^2$} \\ [20pt]
\multicolumn{1}{c}{} &
\multicolumn{1}{l}{$\displaystyle \text{with}\;\; P=P_{3,4}/P_5 \;\; \text{and} \;\; \lambda=\rho_1/\rho_5$} \\ [5pt]
\end{tabular}
\caption{The initial (upper left panel) and final (lower left panel) state of the so-called {\it Riemann} problem, for the case of no relative motion between the two sides ($u_1=u_5=0$). The solid lines mark the pressure $P$, the dashed-dotted lines the density $\rho$, and the dotted line the velocity $v$. Kindly provided by Ewald M\"uller. Right side lists the {\it Rankine--Hugoniot} conditions and the solution to this simplified {\it Riemann} problem.
\label{figs:riemann}}
\end{figure}

The reconstructed quantities are used to calculate the left- and right-hand side values at the cell boundaries. These values are then  exploited as initial conditions to solve the so-called {\it Riemann} problem (see right panel of \fref{figs:riemann}), whose solution provides the fluxes of various quantities (e.g., mass, energy, etc.) across the cell borders. Additional constraints are included to avoid oscillations (e.g., the development of new extrema) in these reconstructions, such as the so-called slope limiters, which estimate the maximum slope allowed for the reconstruction. One way is to demand that the total variation among the interfaces does not increase with time. These so-called {\it total variation\index{total variation diminishing (TVD)}\index{TVD, see total variation diminishing}\index{total variation diminishing schemes (TVD)} diminishing} schemes ({\it TVD}; \cite{1983JCoPh..49..357H}) nowadays provide various different slope limiters, as suggested by different authors.

How to solve the general {\it Riemann} problem, i.e. the evolution of a discontinuity initially separating two states, can be found in textbooks (e.g., \cite{1948sfsw.book.....C}). 
Here, we only want to give a brief description of the solution of a shock\index{shock} tube as an example. 
This corresponds to a system where both sides are initially at rest. The left part of \fref{figs:riemann} shows in the initial (upper panel) and the evolved (lower panel) system. The latter can be divided into five regions. The values for regions 1 and 5 are identical to the initial configuration. Region 2 is a rarefaction wave which is determined by the states in regions 1 and 3. 
The solution can be obtained by invoking the general {\it Rankine--Hugoniot} conditions\index{Rankine--Hugoniot conditions}, describing the jump conditions at a discontinuity. They are listed in the right upper part, where we have assumed a coordinate system which moves with the shock velocity $v_s$. By combining such conditions and defining the initial density ratio $\lambda=\rho_1/\rho_5$, it is possible to get the nonlinear, algebraic equation for the pressure ratio $P=P_{3,4}/P_5$, as displayed in the lower right part of \fref{figs:riemann}. Once $P_{3,4}$ is known by solving this equation, the remaining unknowns can be inferred step by step from the four conditions.

Solving the full {\it Riemann} problem in a hydrodynamics code can be expensive and it can severely affect the code performance. Therefore, there are various approximate methods to solve the {\it Riemann} problem, including the so-called {\it ROE method} (e.g., \cite{1999JCoPh.154..284P}), the {\it HLL/HLLE} method (e.g., see \cite{Harten_1983b, Einfeldt_1988, Einfeldt_1991}), and {\it HLLC} (e.g., see \cite{2005JCoPh.203..344L}). A description of all these methods is outside the scope of this review, so we refer the reader to the references given or to textbooks (e.g., \cite{L2002}).

A wide variety of grid codes are used for cosmological applications, including {\it TVD}-based codes\index{TVD} (e.g., \cite{1993ApJ...414....1R} and {\it CosmoMHD} \cite{2006astro.ph.11863L}) and {\it PLM}-based\index{PLM} codes (like {\it ART} \cite{1997ApJS..111...73K, 2002APS..APR.E2004K} and {\it RAMSES} \cite{2002A&A...385..337T}). 
{\it PPM}-based\index{PPM} codes include {\it Zeus} \cite{1992ApJS...80..753S}, {\it ENZO} \cite{1995CoPhC..89..149B}, {\it COSMOS} \cite{2000ApJ...536..122R}, and {\it FLASH} \cite{2000ApJS..131..273F}. 
It is worth mentioning the {\it WENO}-based\index{WENO} code by \cite{2004ApJ...612....1F}.

\begin{figure}
 \centerline{\includegraphics{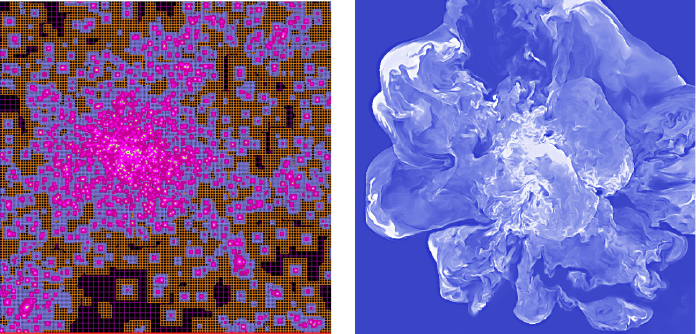}}
 \caption{Left panel: Mesh refinement levels of {\it RAMSES} for a typical, cosmological simulation. Kindly provided by Romain Teyssier. Right panel: Slice in gas temperature through a cluster simulation, with refinement based on velocity jumps. This refinement criterion ensures high resolution around shocks even in the outer parts of the cluster. Figure taken from
\cite{2010NewA...15..695V}.
\label{figs:adaptive_mesh}}
\end{figure}

\section{Adaptive mesh refinement}
To widen the dynamical range of the numerical schemes, mesh refinement strategies have been applied in several grid codes (e.g., {\it ART}, {\it RAMSES}, {\it ENZO}, and {\it FLASH}). In most of the cases, a simple density (e.g., mass per cell) criterion is used. If the mass within one cell exceeds a certain threshold, 
$m \equiv \rho \, \Delta x^3  > m_{\mathrm{min}}$, 
the cell is divided in multiple (e.g., eight) sub-cells and the internal properties are interpolated from the original cell onto the new sub-cells. This ensures that the gravitational mass (e.g., the source of gravity) is homogeneously distributed within the computational domain. In this way, the underlying grid evolves in a quasi-Lagrangian fashion following the mass flow, as illustrated in the left panel of \fref{figs:adaptive_mesh}, which shows the typical structure of the refinement grid in a cosmological simulation.

Other refinement strategies based on velocity criteria are often used \cite{2010NewA...15..695V, 2011ApJ...726...17P} to study shocks and turbulence\index{turbulence} in galaxy clusters\index{galaxy clusters}. 
By extending the aforementioned criteria to additionally refining on velocity jumps, the formation of turbulence and shocks can be followed with unprecedented high spatial resolution throughout the cosmic structures, as shown in the right part of \fref{figs:adaptive_mesh}. 
In order to follow the turbulent cascade with high precision and accuracy, sub-scale turbulence models can additionally be used to initialize the velocities on the refined cells: this prevents the turbulent cascade from being suppressed \cite{2011MNRAS.414.2297I}.

\begin{figure}
\begin{tabular}{m{0.5\textwidth}m{0.5\textwidth}}
\multicolumn{2}{c}{Kernel density estimate} \\ [5pt]
\multicolumn{1}{l}{$\displaystyle \langle X({\vec x}) \rangle = \int\!\!W({\vec x} - {\vec x}',h) X({\vec x}') d {\vec x}'$} &
\multicolumn{1}{l}{\multirow{4}{*}{\includegraphics[width=0.46\textwidth]{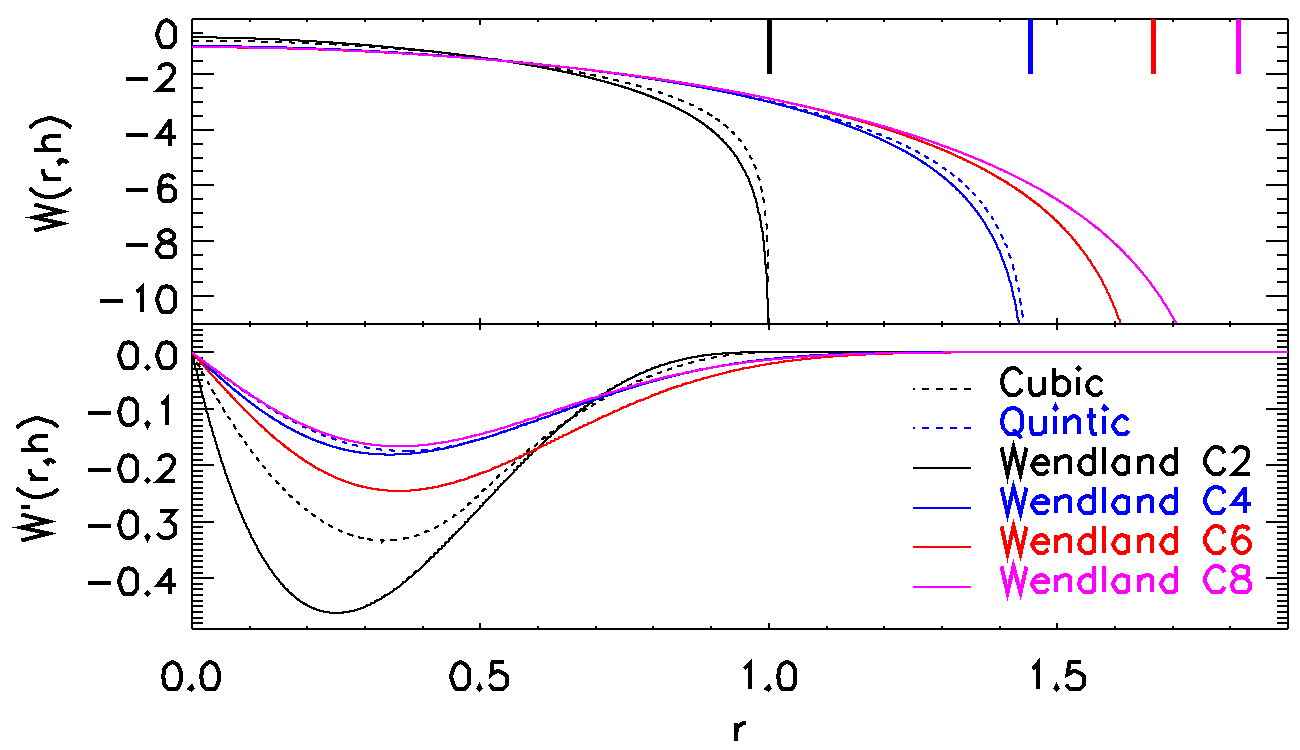}} } \\ [8pt]
\multicolumn{1}{l}{$\displaystyle \int\!\!W({\vec x},h) \dd {\vec x} = 1\;\text{and}\;W({\vec x},h)\!\underset{h \rightarrow 0}{\rightarrow}\!\delta({\vec x})$} &
\multicolumn{1}{c}{} \\ [8pt]
\multicolumn{1}{l}{$\displaystyle \langle X_i\rangle = \langle X({\vec x}_i) \rangle = \sum_j \frac{m_j}{\rho_j} X_j W({\vec r}_{ij},h)$} &
\multicolumn{1}{c}{} \\ [8pt]
\multicolumn{1}{l}{$\displaystyle \nabla_i \langle X({\vec x}_i) \rangle = \sum_j \frac{m_j}{\rho_j} X_j \nabla_i W({\vec r}_{ij},h)$} &
\multicolumn{1}{c}{} \\ [15pt]
\hline \\[-10pt]
\multicolumn{1}{l}{Pairwise symmetric formulation} & \multicolumn{1}{r}{individual smoothing length} \\[5pt]
\multicolumn{1}{l}{$\displaystyle W\!(\vec{r}_{ij},h_i,h_j) = \left\{ \begin{array}{l} \!\langle W\!(\vec{r}_{ij},h_i), W\!(\vec{r}_{ij},h_j) \rangle  \\ W\!(\vec{r}_{ij},\langle h_i,h_j\rangle ) \\ \end{array} \right. $} & \multicolumn{1}{r}{$\displaystyle \frac{4\pi}{3}h_i^3\rho_i = Nm_i$}\\ [10pt]
\multicolumn{1}{l}{$\displaystyle (\rho{\vec\nabla})\cdot{}X={\vec\nabla}(\rho\cdot{}X)-\rho\cdot({\vec\nabla}X)$} & \multicolumn{1}{r}{$\displaystyle {\vec\nabla}_i W({\vec r}_{ij},h_i,h_j) = W'_{ij} \frac{\partial h_i}{\partial \rho_i} = W'_{ij} f_i$}\\ [8pt]
\multicolumn{1}{l}{$\displaystyle \frac{{\vec\nabla}\cdot X}{\rho}={\vec\nabla}\frac{X}{\rho}+\frac{X}{\rho^2}{\vec\nabla}\cdot\rho$} & \multicolumn{1}{r}{$\displaystyle f_i \equiv \left(1 + \frac{h_i}{3\rho_i} \frac{\partial \rho_i}{\partial h_i}\right)^{-1}$} \\ [10pt]
\hline \\[-10pt]
\multicolumn{2}{c}{Estimating higher order gradients from expanding $\vec{v}_j^\beta$ around $i$} \\ [5pt]
\multicolumn{2}{c}{$\displaystyle \sum_j m_j(\vec{v}_j - \vec{v}_i)^\beta \vec{\nabla}_i^\alpha W_{ij} = \partial_\gamma \vec{v}_i^\beta \sum_j m_j(\vec{x}_j - \vec{x}_i)^\gamma \vec{\nabla}_i^\alpha W_{ij}$} \\
\multicolumn{1}{l}{Improved estimate of $\partial_\gamma \vec{v}_i^\beta \equiv\;${\bf X}$_{\gamma\beta}$} & \multicolumn{1}{r}{matrix inversion {\bf X}$=${\bf M}$^{-1}${\bf Y}}\\ [5pt]
\multicolumn{1}{l}{{\bf M}$\displaystyle_{\alpha\gamma} \equiv \sum_j m_j(\vec{x}_j - \vec{x}_i)^\gamma \vec{\nabla}_i^\alpha W_{ij}$} & \multicolumn{1}{r}{{\bf Y}$\displaystyle_{\alpha\beta} = \sum_j m_j(\vec{v}_j - \vec{v}_i)^\beta \vec{\nabla}_i^\alpha W_{ij} $}\\ [15pt]
\multicolumn{1}{l}{$\displaystyle \Rightarrow \vec{\nabla}\cdot\vec{v} = \partial_\alpha \vec{v}^\alpha$} & \multicolumn{1}{r}{$\displaystyle \Rightarrow (\vec{\nabla}\times\vec{v}) = \epsilon_{\alpha\beta\gamma}\partial_\alpha \vec{v}^\beta$} \\

\end{tabular}
\caption{Upper left part shows the basic formulation of a kernel density estimate ($\langle X(\vec{x})\rangle$, upper left part), including the properties of the kernel $W(\vec{x},h)$ (see also upper right panel) and the discretization of the derivatives. The middle part lists various ways for a symmetric formulations (left part) and the changes in modern SPH to account for individual smoothing length $h_i$ (right part). The lower row shows how higher order gradients are estimated using a taylor expansion and matrix inversion in very modern SPH formulations. \label{figs:sph_basics}}
\end{figure}

\section{Lagrangian methods and Smoothed Particle Hydrodynamics}\label{sec:sph}

The classical Lagrangian method is the so-called Smoothed Particle\index{Smoothed Particle Hydrodynamics (SPH)}\index{SPH, see Smoothed Particle Hydrodynamics} Hydrodynamics method (SPH; \cite{1977MNRAS.181..375G, 1977AJ.....82.1013L}), which solves the Lagrangian form of the Euler equations (see \fref{fig:HydroMethods}) and can achieve good spatial resolution in high-density regions.

The basic idea of SPH is to discretize the fluid by mass elements (i.e. particles), rather than by volume elements as in Eulerian methods (see \fref{fig:HydroMethods}). It is evident that the mean interparticle distance in collapsed objects is smaller than in underdense regions. The scheme will thus be adaptive in spatial resolution by keeping the mass resolution fixed. For a comprehensive review with rigorous derivations of the SPH formalism, see~\cite{2012JCoPh.231..759P}. 

\fref{figs:sph_basics} summarizes the main characteristics of SPH.
The general definition of a kernel smoothing method is the first step to build continuous fluid quantities for an arbitrary variable $X$. Here, the kernel depends on the distance modulus only, and is in addition required to be monotonic and differentiable (see upper part of \fref{figs:sph_basics}). In the discretized version, we replace the volume element of the integration, $d {\vec x} = d^3 x$, with the ratio of the mass and density $m_j/\rho_j$ of the particles. Although this equation holds for any position ${\vec x}$ in space, we are only interested here in the fluid representation at the original particle positions ${\vec x}_i$, which are the only locations where we will need the fluid representation later on. It is important to note that for kernels with compact support (i.e., $W({\vec x},h)=0$ for $|{\vec x}|>h$), the summation does not have to be done over all the particles but only over the particles within the sphere of radius $h$ (see \fref{fig:HydroMethods}), namely, the neighbors around the particle $i$ under consideration. Originally, the most frequently used kernel is the $B_2$-Spline, but modern schemes invoke an entire new family of kernels like the HOCT kernels \cite{2010MNRAS.405.1513R} or the so-called Wendland kernels \cite{2012MNRAS.425.1068D} which show better stability and higher accuracy (see upper right panel of \fref{figs:sph_basics} and also recent review by \cite{2015LRCA....1....1R} and references therein).

Derivatives can be calculated using the analytically known derivatives of the kernel. Conservation laws are numerically better achieved when pairwise symmetric formulations are used: this can be obtained by exploiting some identities when using derivatives of $\rho X$ or $X/\rho$ (see left middle part of \fref{figs:sph_basics}). To profit from the adaptive nature of the SPH method, the smoothing\index{smoothing length} length $h_i$ is typically allowed to vary for each individual particle $i$ and is determined by encompassing either a fixed number of neighbours or a fixed mass within the kernel. However, two complications often stem from the abovementioned individual smoothing lengths: first, a pairwise symmetric formulation needs to define an averaged kernel for each particle pair, and historically various ways to build averages have been discussed. The other is that the derivatives of the kernel -- being the smoothing length spatial dependent -- come with correction terms ($\partial h_i/\partial \rho_i$), which originally have been always ignored because they can't be computed directly. Starting from an entropy\index{entropy formulation} formulation, \cite{springel02} derived for the first time an SPH formulation including the proper correction terms for the varying smoothing length from a Lagrangian formalism, which are listed in the middle right part of \fref{figs:sph_basics}. Note that this formalism then also specifies the way in which the averaged kernel for a particle pair has to be constructed. 

This can be further generalized, as shown in \cite{2013MNRAS.428.2840H}. There, such a generalized formulation was obtained starting from an x-weighted volume average
\begin{equation}
\bar{y} = y_i = \sum_j x_j W_{ij}(h_i),
\end{equation}
the (x-weighted) volume element $\Delta\nu_i\equiv x_i/y_i$ and the generalized relation
\begin{equation}
P_i = (\gamma-1)u_i\frac{m_i}{\Delta\nu_i} = A_i\left(\frac{m_i}{\Delta\nu_i}\right)^\gamma
\end{equation}
between the pressure $P_i$, the internal energy per unit mass $u_i$ and the entropic function $A_i$. This leads to the set of generalized SPH equations
\begin{equation}
m_i \frac{d {\vec v}_i}{d t} = - \sum_j x_i x_j \left(f_{ij}\frac{P_j}{y_j^2}{\vec \nabla}_i W_{ij}(h_i) + f_{ji}\frac{P_i}{y_i^2}{\vec \nabla}_i W_{ij}(h_j) \right)\!,
\end{equation}
and
\begin{equation}
f_{ij} \equiv 1 - \frac{\tilde{x_i}}{x_j} \left(\frac{h_i}{3\tilde{y_i}} \frac{\partial y_i}{\partial h_i}\right) \left[1 + \frac{h_i}{3\tilde{y_i}}\frac{\partial \tilde{y_i}}{\partial h_i}\right]^{-1}.
\end{equation}
For the choice of $x_i=\tilde{x_i}=m_i$, which implies $y_i=\tilde{y_i}=\bar{\rho_i}$ and $\Delta\nu_i=m_i/\rho_i$, and following the entropy $A_i$ (e.g., $P_i=A_i\bar{\rho_i}^\gamma$), this set of equations will result in the entropy-conserving formulation of SPH as presented in \cite{springel02}. For the choice of $x_i=\tilde{x_i}=(\gamma-1)m_i u_i$, implying $y_i=\bar{P_i}$ and $\Delta\nu_i=(\gamma-1)m_i u_i/P_i$, this results in a pressure--energy formulation of SPH, as presented in \cite{2013ApJ...768...44S}. As now pressure is a kernel weighted quantity, contact discontinuities are properly treated. A third possibility is to choose $x_i=m_iA_i^{1/\gamma}$ which leads to a pressure--entropy formulation. For more
details, see \cite{2013MNRAS.428.2840H}.

The SPH method, however, is not performing as well in low-density regions as in the higher density ones. It also suffers from degraded resolution in shocked regions due to the introduction of a sizeable\index{artificial viscosity} artificial viscosity. In its classical implementation, discretization errors introduce spurious pressure forces on particles in regions with steep density gradients in particular near contact discontinuities. This results in a boundary gap of the size of an SPH smoothing kernel radius over which interactions are severely damped. The standard implementation typically does not involve an explicit mixing term, which can compensate this effect. Therefore, the classical implementation does not resolve and treat dynamical instabilities in the interaction of multi-phase fluids, such\index{Kelvin--Helmholtz} as {\it Kelvin--Helmholtz} or {\it Rayleigh--Taylor} instabilities\index{Rayleigh--Taylor instabilities}. Both these shortcomings can be overcome by a modern implementation as described at the end of this section. In addition, in the cosmological context, the adaptive nature of the SPH method, its simple way to couple to gravity and the possibility to have individual time steps often compensate for such shortcomings, thus making SPH still one of the most commonly used methods in numerical hydrodynamical cosmology.

As mentioned before, we also need to add a so-called {\it artificial\index{artificial viscosity} viscosity}, $\Pi_{ij}$, so that he final equations for the entropy-conserving formulation of SPH are reading

\begin{equation}
\frac{d {\vec v}_i}{d t} = - \sum_j m_j \left(f_j\frac{P_j}{\rho_j^2}{\vec \nabla}_i W_{ij}(h_j)+f_i\frac{P_i}{\rho_i^2}{\vec \nabla}_i W_{ij}(h_i) +
\Pi_{ij} {\vec \nabla}_i {\bar W}_{ij} \right),
\end{equation}
and
\begin{equation}
\frac{d A_i}{d t} = \frac{1}{2} \frac{\gamma-1}{\rho_i^{\gamma-1}} \sum_j m_j \Pi_{ij}
\left({\vec v}_j - {\vec v}_i\right){\vec \nabla}_i {\bar W}_{ij}.
\end{equation}

This term describing an {\it artificial viscosity} is usually needed to capture shocks\index{shock} and its construction is similar to other hydrodynamical schemes. Popular formulations are those proposed by Monaghan and Gingold \cite{1983JCoPh..52....374S} and Balsara~\cite{1995JCoPh.121..357B}, which includes a bulk viscosity and a {\it von~Neumann--Richtmeyer} viscosity term, supplemented by a limiter reducing angular momentum transport in the presence of shear flows at low particle numbers \cite{1996IAUS..171..259S}. Modern schemes implement a form of the artificial viscosity as proposed by~\cite{1997JCoPh..136....298S}, based on an analogy with {\it Riemann} solutions of compressible gas dynamics. To reduce this artificial viscosity, at least in those parts of the flows where there are no shocks, a posibility is to follow the idea proposed by Morris and Monaghan~\cite{1997JCoPh..136....41S}: every particle carries its own artificial viscosity, which eventually decays outside the regions which undergo shocks. A detailed study of the implications on the ICM of such an implementation can be found in \cite{2005MNRAS.364..753D}. There are various further improvements on the implementation a higher order artificial dissipation term \cite{2010MNRAS.408..669C,2012MNRAS.422.3037R}. Modern SPH formulations also make use of higher order calculation schemes for velocity gradients, see lower part of \fref{figs:sph_basics} as well as related discussion in \cite{2012MNRAS.420L..33P,Hu2014}.  Even better suppression of the artificial viscosity\index{artificial viscosity} can be reached by following \cite{2010MNRAS.408..669C} in combination with such higher order calculation schemes for velocity gradients, as shown in \cite{Beck2016}.

Considerable effort has been made in the last decade to combine the advantages of Langrangian and Eulerian methods. On one hand, significant progress has been achieved in involving Godunov methods\index{Godunov methods} into the SPH formalism, see \cite{2002JCoPh.179..238I,2003MNRAS.340...73C,2010MNRAS.403.1165C,2011MNRAS.417..136M}, which recently led to the so called meshless methods. On the other hand, large effort has been made to extend Eulerian methods to moving mesh methods. Both these new methods are briefely described in the next two sub-sections.

\section{Moving\index{Moving MESH} mesh methods}
\begin{figure}
 \centerline{\includegraphics{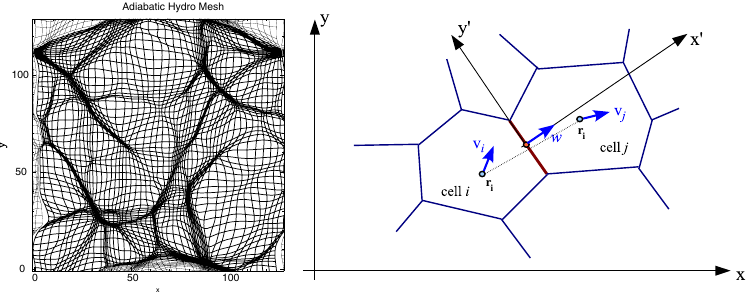}}
 \caption{Left panel: A layer through a cosmological, $128^3$ moving mesh simulation. Figure taken from \cite{1998ApJS..115...19P}. Right panel: Illustrating the concept of flux computation on a Voronoi mesh. Figure taken (and slightly adapted) from \cite{2010MNRAS.401..791S}.
\label{figs:moving_mesh}}
\end{figure}

Substantial effort has gone into reformulating Eulerian methods as described in Section~\ref{grid} into Lagrangian mesh approaches. More details on the idea of hydrodynamics on moving mesh for cosmological application can be found in the pioneering work by \cite{1998ApJS..115...19P} and references therein. This early approach started from a regular mesh which then, by following the flow of the fluid, was deformed. The Euler equations were evolved by calculating the fluxes across the cell borders. An example of the resulting mesh for a cosmological simulation can be seen in the left part of \fref{figs:moving_mesh}. One disadvantage (or challenge) of this technique in practical applications is that individual cells can be extensively deformed and stretched. Modern schemes circumvent this problem by constructing an unstructured mesh based on a Voronoi or Delaunay tessellation\index{Voronoi or Delaunay tessellation} (see \cite{2011MNRAS.414..129G} and references therein). The relevant geometry of the cells, based on the mesh generating points $\vec{r}_i$ and $\vec{r}_j$ is illustrated in the right part of \fref{figs:moving_mesh}. The fluxes then have to be calculated at the centroid of the interface (note that this is not necessarily on the straight line between the two mesh generating points, as indicated by the dotted line)
\begin{equation}
{\bf Q}_i^{(n+1)} = {\bf Q}_i^{(n)} - \Delta t \sum_j A_{ij}\hat{{\bf F}}_{ij}^{(n+1/2)}.\label{eq:flux}
\end{equation}
The motion $\vec{w}$ of this interface is uniquely defined by the velocities $\vec{w}_i$ and $\vec{w}_j$, and the fluxes have to be calculated with the Riemann solver in the rotated frame ($x',y'$). A detailed description of this technique along with its performance in test problems can be found in \cite{2011MNRAS.414..129G}.

\section{Meshless methods}
A new class of Lagrangian methods, the so-called meshless formulations\index{meshless formulations}, have been recently developed for astrophysical problems. More details can be found in \cite{2011MNRAS.414..129G,2015MNRAS.450...53H,Groth2023}, which follow earlier, pioneering work by~\cite{Vila1999,LansonVila2008a,LansonVila2008b}. In short, the derivation starts from the integral form
\begin{equation}
\int[u(\vec{x},t)\dot\phi(\vec{x},t) + \vec{F}(u,\vec{x},t)\cdot\nabla\phi(\vec{x},t) + S(\vec{x},t)\phi(\vec{x},t)]\,d\vec{x}\,dt = 0
\end{equation}
of a scalar conservation law
\begin{equation}
\frac{\partial u}{\partial t} + \nabla\cdot(\vec{F} + \vec{a}u) = S \,.
\end{equation}
Here, $u(\vec{x},t)$ is a scalar field, $S(\vec{x}, t)$ is its source, $\vec{F}(u,\vec{x},t)$ is its flux in a frame moving with velocity $\vec{a}(\vec{x}, t)$, and $\phi(\vec{x},t)$ is an arbitrary differentiable function in space and time leading to the advective derivative $\dot\phi(\vec{x}, t) = \partial\phi(\vec{x}, t)/\partial t + \vec{a}(x,t)\cdot\nabla\phi(\vec{x}, t)$.

\begin{figure}
 \centerline{\includegraphics{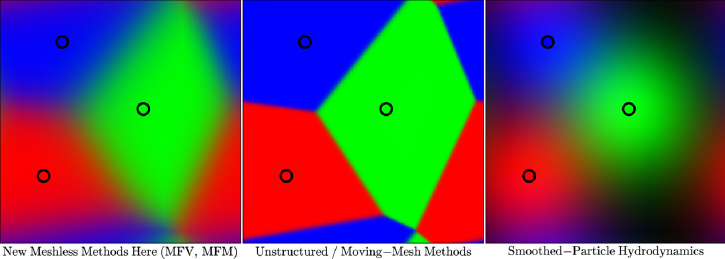}}
 \caption{Illustrating the conceptual differences in partitioning the volumes between meshless methods, unstructured grid and classical kernel weighted formalism, as used in SPH methods. Figure taken from \cite{2015MNRAS.450...53H}.\label{figs:hopkins}}
\end{figure}

Using a discretization through a set of particles $i$ with a smoothing length $h(\vec{x})$ and a kernel function $W(\vec{x},h)$, akin to what is done in SPH, the partitioning of the particles can be written as
\begin{equation}
\psi_i(\vec{x}) = w(\vec{x})W(\vec{x}-\vec{x}_i,h(\vec{x})),
\end{equation}
where the number density of particles is $w(\vec{x})^{-1} = \sum_j W(\vec{x}-\vec{x}_j,h(\vec{x}))$. The discretization of an arbitrary function $f(\vec{x})$ can be written as
\begin{equation}
\int f(\vec{x})\,d\vec{x}  \approx \sum_i f_i \int \psi_i(\vec{x})\,d\vec{x} \equiv \sum_i f_i
V_i,\label{eq:discretisation}
\end{equation}
where $V_i = \int \psi_i(\vec{x})\,d\vec{x}$ is the effective volume of a particle $i$. Therefore, the discrete form of the integral equation (3.28) can be written as
\begin{equation}
\sum_i \int [ V_i u_i \dot\phi_i + V_i F^\alpha_i (D^\alpha\varphi)_i + V_i S_i \phi_i ] = 0 \, .
\end{equation}
Although in principle an SPH estimate for $(D^\alpha\phi)_i$ could be used, it is convenient to use a more accurate meshless gradient estimate, as suggested by \cite{LansonVila2008a}. This, along with the integration of the first term by parts, makes it possible (as shown in \cite{2011MNRAS.414..129G}) to split $\phi(\vec{x}, t)$ and to obtain:
\begin{equation}
\frac{d}{dt}(V_i u_i) + \sum_j\left[V_i F_i^\alpha\psi_j^\alpha(\vec{x}_i) - V_j F_j^\alpha\psi_i^\alpha(\vec{x}_j)\right] = V_i S_i \, .
\end{equation}
This equation (and its extension to a general vector field $\vec{u}$) is very similar to the finite volume equation for the moving mesh method\footnote{Note that, however, \eref{eq:flux} is in the integral form.} but also somewhat similar to the SPH equations, except that the interactions between different particles is described in the source and flux terms, which can be obtained as the solution of an approximate Riemann problem between particles $i$ and $j$. A closer inspection also reveals that only the projection on the direction between the particles is needed, e.g., the solution of the Riemann problem at the midpoint\footnote{More accurate would be the quadrature point at an equal fraction of the kernel length $h_i$ and $h_j$, see discussion in \cite{2015MNRAS.450...53H} and references therein.}. This also means that the primitive variables have to be extrapolated to the midpoint, e.g., using linear extrapolation, and generally a flux limiter has to be applied (see discussion in \cite{2011MNRAS.414..129G,2015MNRAS.450...53H,Groth2023}). \Fref{figs:hopkins} illustrates the differences in partitioning the volumes between mesh-less methods, unstructured (moving) grid and classical kernel weighted \hbox{formalism} (SPH).

\section{Code comparison in galaxy and cluster simulations}
The Eulerian and Lagrangian approaches described in the previous sections are theoretically supposed to provide the same results when applied to the same problem. To verify that codes succeed at correctly integrating the set of hydrodynamical equations, they are usually tested against problems whose solution is known analytically. In practice, these test problems are shock tubes or spherical collapse problems.

However, idealized hydrodynamical tests like the interaction of multi-phase fluids \cite{2007MNRAS.380..963A} often reveal fundamental differences among results obtained with different methods. Such differences can be driven by the formulation of the underlying fundamental equations (like no mixing in classical SPH), by the discretization (like the volume bias in SPH formulations) or they can be due to the influence of numerical errors (like the departure of translation invariance in grid codes due to errors in the reconstruction).

In cosmology, problems with known analytical solutions are impractical: as a consequence, a meaningful test is represented by  comparing the results provided by different codes when they simulate the formation of cosmic structures in a standard set-up. As an example, \cite{2005ApJS..160....1O} compares the thermodynamical properties of the IGM\index{IGM} predicted by the {\it GADGET} (SPH-based) and {\it ENZO} (grid-based) codes. Another example of a comparison between grid-based and SPH-based codes can be found in \cite{1994ApJ...430...83K}.

A detailed comparison of hydrodynamical codes which simulate the formation and evolution of a galaxy or of a galaxy cluster is therefore an extremely important test. A pioneering comparison was performed within the so-called Santa Barbara Cluster Comparison\index{Santa Barbara Cluster Comparison Project} Project \cite{1999ApJ...525..554F}. Here, 12 different groups, each using a code either based on the SPH technique (seven groups) or on the grid technique (five groups), performed a non-radiative simulation of a galaxy cluster from the same initial conditions. 

\begin{figure}
 \centerline{\includegraphics[scale=0.3]{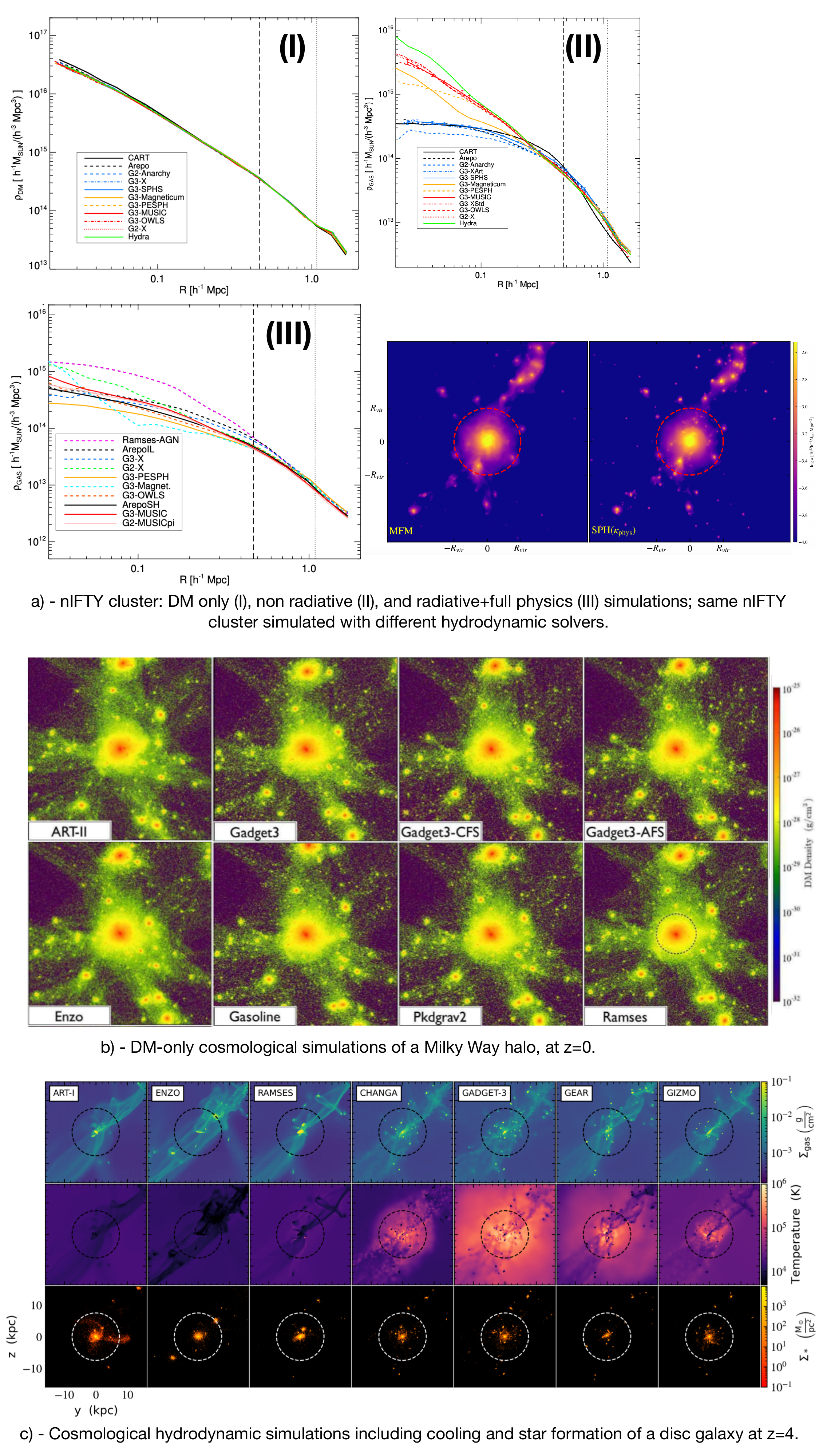}}
 \caption{Selection of figures from code comparison campaigns that aimed at quantifying the impact of using different codes on final results.
 {\it{Panel a:}} Impact of different codes on the density radial profiles of a galaxy cluster, when different simulation setups are considered. We show DM-only simulations $(i)$, non-radiative runs $(ii)$, and radiative plus full physics simulations $(iii)$ (Figures 2 and 6 from \cite{2016MNRAS.457.4063S}, 
 and Figure 8 from \cite{2016MNRAS.459.2973S}). 
 The fourth composite panel shows the same nIFTY cluster simulated with two different hydro solvers (namely, MFM on the left and SPH on the right) by \cite{Groth2023}.
 {\it{Panel b:}} DM-only cosmological simulations of a Milky Way halo, at z=0 (adapted from Figure 3 in \cite{Kim2014_Agora}).
 {\it{Panel c:}} Cosmological hydrodynamical simulations featuring cooling and star formation, and targeting a Milky-Way like galaxy, at z=4 (Figure 11 from \cite{Agora2021}).
 \label{figs:nifty}}
\end{figure}

A more recent comparison project, the so-called nIFTy galaxy cluster\index{nIFTy galaxy cluster simulations} simulations \cite{2016MNRAS.457.4063S}, also involved modern SPH implementations as well as the moving mesh code {\it AREPO}. 
A similar agreement over large ranges was obtained for many of the gas properties in both studies, like the density, temperature and entropy\index{entropy} profiles, as shown in \fref{figs:nifty}. 
Both studies found significantly larger differences between mesh-based codes and classical particle-based codes to be present for the inner part of the profiles. However, as shown in \cite{2016MNRAS.457.4063S}, particle-based codes which include an explicit treatment of mixing produce results very similar to grid-based codes.

A few code comparison campaigns have targeted galaxies as well (see e.g., \cite{Scannapieco2012, Agora2021}, \fref{figs:nifty}, and Section~\ref{sfr}).

It is worth mentioning that as soon as additional physics like star formation and AGN feedback is included, the discrepancies in the results obtained by the various numerical methods are no longer driven by differences which can be ascribed to the hydrodynamical solvers: they are rather due to the details of the numerical prescriptions adopted to model these additional processes \cite{2016MNRAS.459.2973S}.

\section{Sub-resolution modelling of astrophysical processes}
The astrophysical processes involved in the formation and evolution of galaxies in cosmological hydrodynamical simulations span a huge dynamical range of scales. They indeed vary from the $\sim$Mpc scales where gravitational instabilities drive the evolution of the DM component and the hierarchical assembly of haloes, down to the $\sim$parsec and $\sim$sub-parsec scales, where processes like BH accretion and star formation take place, going through the $\sim$kpc scales, where e.g. galactic winds distribute the stellar feedback energy to the ambient medium. 
The call for sub-grid physics is thus essential: sub-resolution models account for processes that occur below the resolution limit of cosmological hydrodynamical simulations, but that affect the evolution of the simulated structure on scales that are explicitly resolved. Sub-resolution prescriptions usually resort to rather simple analytical or theoretical models, and/or to the phenomenological description of rather complex processes, through a suitable choice of parameters that are calibrated to reproduce observations (see also sub-section~\ref{Ch3_agn_feedback}). 

A common criticism that has been often raised to numerical astrophysics deals with the number of parameters that are employed when modelling physical processes, in particular when referring to the parameters that enter sub-resolution models. However, observed or expected results cannot be reproduced by just fine-tuning parameters. Results are rather determined by proper parametrizations of processes. For instance: when modelling star formation (see sub-section~\ref{SF}), results are mainly driven by the way in which the star formation rate (SFR) relates to the actual fuel (e.g. the mass of molecular or cold gas available), rather than to the exact value of the proportionality constant between the two aforementioned quantities. 
Parameter space exploration is thus not meant to provide the best fit to observations that simulations want to compare with. It rather aims at enabling a better understanding of the dependence of a model on different unknowns and at making predictions for loosely constrained physical quantities.

In the next sub-sections, we will review the most important astrophysical processes that significantly affect galaxy formation and evolution. They are all self-consistently included in state-of-the-art cosmological hydrodynamical simulations.

\section{Gas\index{cooling} cooling}\label{cooling}
Hot gas cooling enters the right-hand side of equation~\ref{equation:firstlaw}, as well as the equations describing mass and energy flows among different gas phases which may be present in sub-resolution models.

\begin{figure}
 \centerline{\includegraphics{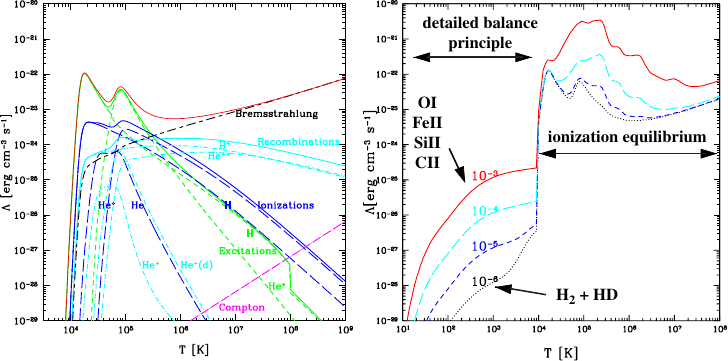}}
 \caption{Left panel: The total cooling curve (solid line) and its composition from different processes for a primordial mixture of H and He. Figure taken from \cite{Katz1996}. Right panel: The total cooling curve as a function of different metallicity. The part below $10^4$~K also takes into account cooling by molecules (e.g., HD and H$_2$) and\index{metal lines} metal lines. Figure taken from \cite{Maio2007}.
 \label{figs:cooling}}
\end{figure}

In cosmological simulations, the focus is usually on structures whose virial temperature exceeds $\sim 10^4$~K. Common assumptions in standard implementations of the cooling\index{cooling function} function $\Lambda(u,\rho)$ is that the gas is optically thin and in ionization equilibrium. 
It is also usually assumed that three-body cooling processes are unimportant so as to restrict the treatment to two-body processes. For a plasma with primordial composition of H and He, these processes are: \index{collisional excitation} collisional excitation of H$^0$ and He$^+$, collisional ionization of H$^0$, He$^0$, and He$^{+}$, standard recombination\index{recombination} of H$^+$, He$^+$, and He$^{++}$, dielectric recombination of He$^+$, and free--free\index{bremsstrahlung} emission (bremsstrahlung). 
The collisional ionization and recombination rates depend only on temperature. Therefore, should an \index{ionizing background} ionizing background radiation be absent, the resulting rate equations can be solved analytically. This leads to a cooling function $\Lambda(u)/\rho^2$ as illustrated in the left panel of \fref{figs:cooling}. 
On the other hand, in the presence of ionizing background radiation, the rate equations can be solved iteratively. Note that for a typical cosmological radiation background (e.g., UV background from star-forming galaxies and quasars, see \cite{HaardtMadau1996, HaardtMadau2012, Puchwein2019, Faucher-Giguere2020}), the shape of the cooling function can be significantly altered, especially at low densities. For a more detailed discussion, see, for example, \cref{Ch_galform} and \cite{Katz1996}.

Additionally, the presence of metals will drastically increase the number of possible processes by which gas can cool. As it becomes computationally very demanding to calculate the cooling function in this case, cosmological simulations usually rely on a pre-computed, tabulated cooling function. As an example, the right panel of \fref{figs:cooling} shows the tabulated cooling function from \cite{1993ApJS...88..253S}, for temperatures above $10^5$~K, for different metallicities of the gas, keeping the ratios of different metal species fixed to solar values. 
A refinement that is nowadays often exploited in simulations consists in using the {\tt Cloudy} code (\cite{ferland98}), which allows users to tabulate the cooling and heating rates, individually for a grid of UV intensities and for various different chemical elements. In this way, cooling rates are self-consistently calculated for arbitrary chemical
compositions. 
Note that almost all the implementations solve the above rate equations (and therefore the cooling\index{cooling} of the gas) as a ``subtime step'' problem, decoupled from the hydrodynamical treatment. This translates in assuming that the density is fixed across the time step. 
Furthermore, the time step of the underlying hydrodynamical simulation is in general, for practical reasons, not controlled by nor related to the cooling time scale. The resulting uncertainties introduced by these approximations have not yet been deeply explored and clearly leave room for future investigations.

For the formation of the first objects in halos with virial temperatures\index{virial temperature} below $10^4$~K, the assumption of ionization equilibrium no longer holds. In this case, the non-equilibrium reactions have to be considered, by solving the balance equations for the individual levels of each species during the cosmological evolution. 
In the absence of metals, the main coolants are H$_2$ and H$_2^+$ molecules (see \cite{1997NewA....2..181A}). HD molecules can also play a significant role. When metals are present, many more reactions are available and some of these can contribute significantly to the cooling\index{cooling function} function below $10^4$~K. This effect is clearly visible in the right panel of \fref{figs:cooling} for $T<10^4$~K. For more details, see \cref{Ch_galform}, Refs.~\cite{1998A&A...335..403G, Maio2007}, and references therein.

\section{Star formation}\label{SF}

Star formation is a key astrophysical process which involves a hierarchy of physical processes: the accretion of gas from the large ($\sim$Mpc) scale, gas cooling to produce neutral (HI) and molecular (H$_{\rm 2}$) hydrogen ($\sim$kpc scale), the formation of giant molecular clouds (GMCs; $\sim$10$\div$100~pc), H$_{\rm 2}$ fragmentation and accretion to clumps ($\sim$1~pc) and cores ($\sim$0.1~pc), and the subsequent contraction of cores to form stars ($\sim$10$^{-8}$~pc). These complex processes, along with the description of the properties of the multi-phase star-forming ISM and the different regimes for star formation are thoroughly reviewed in a number of works, e.g. \cite{McKee_Ostriker_2007, Kennicutt_Evans_2012, Madau_Dickinson_2014, Tacconi_2020}. 

Star formation is the outcome of the gravitational collapse of clouds of dense and cold gas. Being the sub-pc scale where star formation actually occurs not captured in state-of-the-art cosmological hydrodynamical simulations, this is a sub-resolution process. From an order-of-magnitude estimate of masses and scales involved in cosmological simulation, gas elements have usually masses in the range of $10^4 \div 10^7$~M$_{\odot}$ (they do not resolve GMCs), and have associated softenings comparable with or larger than a few hundreds pc (depending on resolution). Star particles have mass and force resolution comparable to gas elements, and are treated as SSPs. In summary, star formation in cosmological simulations aims at effectively capturing the drop out of cold gas which underwent radiative cooling into collisionless stellar particles.

A gas particle or cell is eligible to form stars if a number of criteria are met: the flow has to be locally convergent ($\nabla \cdot \mathbf{v} < 0$) and Jeans unstable (i.e. the sound-crossing time exceeds the dynamical time); gas overdensity is larger than a critical overdensity (to prevent star formation in low-overdensity gas at very-high redshift); and the local number density of hydrogen atoms exceeds a threshold number density (see below). Details about the aforementioned conditions are provided in \cref{Ch_galform} (see also \cite{Katz1996}).

A seminal work which advanced the description of the star-forming ISM in cosmological simulations and the numerical modelling of star formation is \cite{Springel2003}. They introduced a model for the multiphase ISM, that accounts for star formation and stellar feedback. It considers that hot and cold gas components coexist in pressure equilibrium within dense gas particles, that are the so-called multiphase particles. Properties of the cold and hot phases are assumed to represent averages over small volumes of the ISM, since molecular clouds cannot be resolved. The cold phase within the multiphase particle provides the reservoir for star formation, while being replenished by hot gas cooling. A fraction of the cold gas fuelling star formation is assumed to be instantaneously used to heat up the hot phase and to evaporate a share of the cold phase, mimicking the explosion of short-lived massive stars as SNe. Equations describing the evolution of the energy of the hot and cold gas phases are integrated, so that the star formation rate can be computed. Parameters of the model are calibrated to reproduce the Schmidt-Kennicutt relation \cite{schmidt59,kennicutt1989}. 
This model assumes a constant temperature ($\sim 10^3$~K) for the cold gas component. Also, it is characterised by a quiescent, self-regulated star formation, as the fuelling of the cold phase owing to cooling is counterbalanced by its evaporation due to SN heating. Within this model, the ISM can be described by an effective equation of state: the effective pressure of the two-phase gas particles has a functional form that is constrained by the properties of the hot and cold phases, and is expected to be constant in time due to the self-regulated star formation. 

Several sub-resolution models widely adopted in state-of-the-art cosmological simulations are inspired or loosely based on the aforementioned model. 
An interesting variation of this model considers that the density threshold for gas particles to become eligible to form stars depends on metallicity, based on the assumption that the higher the metallicity of the gas, the lower the density for warm and neutral gas to become cold and molecular \cite{Schaye2004, Schaye2015}. 

A relevant feature of the Springel and Hernquist model is the use of the effective equation of state for star-forming gas. It stems from the idea that the small-scale effects (e.g. thermal instabilities, turbulence, thermal conduction) which regulate the ISM are responsible for setting a self-regulated, equilibrium state of the ISM over short timescales. Within a similar regime, the average ISM temperature or energy can be approximated as a function of the warm/cold gas density only (see \cite{Valentini2017, Marinacci2019} and \cite{Wang2015} for examples of simulations not relying on the equation of state or on the presence of multiphase resolution elements, respectively).

As discussed above, the majority of numerical simulations of structure formation in a cosmological context are still far from resolving the structure of GMCs and from modelling star formation from first principles. They rather rely on various sub-resolution prescriptions that connect the SFR with the available gas eligible to form stars. This often translates in considering cold and dense gas ($T \lesssim \mbox{a few} \times 10^4$~K; $n\gtrsim 0.1$-$10$ cm$^{-3}$), and in the assumption that the local SFR is proportional to its density over a timescale (close to the dynamical time of the cold gas). The aforementioned density and temperature thresholds to select star-forming gas actually reflect the physical properties of HI in the ISM. The latter proportionality is loosely inspired by the Schimdt-Kennicutt law \cite{schmidt59,kennicutt1989}.

Examples of state-of-the art cosmological simulations that adopt this approach feature Illustris \cite{Vogelsberger2014}, Magneticum \cite{Dolag2015}, NIHAO \cite{Wang2015}, Illustris-TNG \cite{Pillepich2018}, the Massive-Black \cite{Khandai2015} and Fable \cite{Henden2018} simulations, the Horizon-AGN suite \cite{Dubois2016, Dubois2021}, the DIANOGA simulations \cite{ragone18,Bassini2020}, the SLOW simulations \cite{Dolag2023}.

Among the cosmological simulations which retrieve the SFR from the availability of molecular gas, estimated through the theoretical model by \cite{KrumholzGnedin2011}, it is worth mentioning the MUFASA and SIMBA simulations \cite{Dave2017, Dave2019}, the FIRE suite \cite{Hopkins2014, Hopkins2017} and FIREBOX \cite{Feldmann2023}.

Other examples of cosmological simulations with a sub-resolution treatment of H$_{\rm 2}$ include e.g., \cite{pelupessy06, Maio2007, Feldmann2011, Christensen2012, Kuhlen2012, Murante2015, Tomassetti2015, Lupi2018, Pallottini2019, Schaebe2020, Valentini2023}. Targeting the aforementioned simulations single haloes or smaller volumes, they manage to achieve a more refined modelling of the star-forming, molecular gas.

In what follows, we provide a short overview of the different prescriptions adopted by state-of-the art cosmological simulations to select star-forming gas and to model star formation. The below-listed conditions usually come in addition to the ones mentioned at the beginning of this sub-section.
The Illustris, Magneticum, Illustris-TNG and SLOW simulations all select the star-forming gas based on a density threshold ($n_{\ast} \simeq 0.1$~cm$^{-3}$), and assume a constant depletion timescale to convert cold gas into stars and estimate the SFR (based on \cite{Springel2003}).
The NIHAO simulations and the ERIS and GIGA-ERIS simulations \cite{Guedes2011, Tamfal2022} adopt both a density and a temperature threshold to select star-forming gas. 

Density and temperature thresholds are adopted also by simulations which rely on the MUPPI sub-resolution model (e.g. \cite{Murante2010, Murante2015, Valentini2017, Valentini2019, Granato2021, Parente2022, Valentini2023}) to select multi-phase gas: star formation then proceeds from the fraction of cold gas which is molecular, as estimated via either the \cite{Blitz2006} or the \cite{Krumholz2009} prescription; finally, the depletion timescale is the dynamical time of the cold gas. A more advanced modeling within MUPPI profits from the treatment of dust formation and evolution \cite{Granato2021, Parente2022} to predict the evolution of H$_{\rm 2}$ \cite{Ragone-Figueroa2024}, which in turn is used to estimate the local SFR.

In a similar fashion, the MUFASA and SIMBA simulations adopt an H$_{\rm 2}$-based prescription for the star formation following the \cite{KrumholzGnedin2011} formula (considering that gas whose density exceeds a threshold is star forming), and assume that the depletion timescale is the local dynamical time. Similar assumptions to estimate the molecular gas out of the cold phase are made in the FIRE simulations.

The EAGLE and FLAMINGO simulation suites \cite{Schaye2015, Schaye2023} assume a metallicity-dependent star formation threshold \cite{Schaye2004} along with a temperature threshold: aiming at tying star formation to the molecular gas, they exploit the {\tt Cloudy} radiative transfer code \cite{ferland98} to determine the transition from the warm, atomic to the cold, neutral gas phase. In these simulations, the SFR depends on pressure rather than on density.
Interestingly, the HORIZON simulation suite (e.g., \cite{Dubois2021}) -- which selects star-forming gas from gas cells above a given density threshold -- features a locally-varying star formation efficiency (at variance with almost all the aforementioned simulations, where the fraction of cold/molecular gas which is converted into stellar particles is assumed to be constant).

\cite{Hopkins2013, Buck2019} address how different criteria adopted to select star-forming gas have a strong impact on simulated galaxies.

Star formation is often implemented according to a stochastic model: the process is not described for each star, rather stochastically with expectation values in agreement with the actual SFR. If, e.g., the probability $p = m_{\rm gas}/m_{\star} (1- {\text{exp}}(m_{\star}/m_{\rm gas}))$ exceeds a random number, then a new star particle is spawned \cite{Springel2003}.
As a refinement, a number of stellar generations can be assumed \cite{tornatore07}, so that newly spawned stellar particles have a given share of the mass of the gas element they come from as their initial mass.

\section{Stellar evolution and chemical enrichment}\label{chemical_enrichment}

Chemical evolution is a natural outcome of galaxy evolution in cosmological hydrodynamical simulations. Models of chemical evolution have been included in cosmological simulations of cosmic structure formation to properly address the study of the chemical enrichment process: the distribution of metals in the ISM and in the circum-galactic medium (CGM) encodes valuable information of the past history of star formation and feedback, and is a crucial feature of hydrodynamical simulations (e.g. \cite{Oppenheimer2018, Torrey2018}).
At variance with observations that explicitly resolve individual stars, cosmological simulations provide a coarse sampling of stellar populations by means of star particles. 
The majority of state-of-the-art cosmological hydrodynamical simulations of both large cosmological volumes and individual galaxies have star particles whose mass typically ranges between $10^8$ and $10^3$~M$_{\odot}$ according to resolution (e.g. \cite{Dubois2014, Vogelsberger2014, Khandai2015, Rasia2015, Schaye2015, McCarthy2017, Pillepich2018, Schaye2023, Oser2010, Aumer2013, Stinson2013, Marinacci2014, Murante2015, Valentini2017, Hopkins2018}). 
In these simulations, star particles are resolution elements: each of them represents a simple stellar population (SSP), i.e. an ensemble of coeval stars that share the same initial metallicity. Every stellar particle initially shares the chemical composition of the gas element from which it has been originated, and is characterized by an initial mass function (IMF). 

The IMF\index{initial mass function (IMF)}\index{IMF, see initial mass function} $\phi (m)$ determines the number of stars per unit mass interval: 
\begin{equation}
\phi (m) =  \beta m^{- \alpha} \,\,\,,
\label{IMF}
\end{equation}
within a given mass range $[M_{\rm inf}, M_{\rm sup}]$. The coefficient $\alpha$ sets the slope of the power law over the mass range or in the different mass intervals within the mass range. For each mass interval, a normalization constant $\beta$ is computed by imposing that $\int m \, \phi (m) \, dm = 1$ over the global mass range and continuity at the edges of subsequent mass intervals.

Commonly assumed IMFs consist of single- \cite{Salpeter1955, Arimoto1987} or multiple-slope \cite{Kroupa1993, Kroupa2001, Kroupa2002} power law, but more sophisticated shapes have been proposed (e.g. \cite{Chabrier2003}). We refer the reader to \cite{Borgani2008, Few2014, Bekki2013, GutckeSpringel2017, Barber2018, Biffi2018, Valentini2019} for details about different IMFs and to appreciate how adopting different IMFs can have an impact on metal distribution on cosmological simulations. 

The IMF directly regulates the relative ratio between stars of different initial mass, thus affecting the relative abundance of elements contributed by different types of stars. 
Stellar evolution predicts that all stars with masses larger than $M_{\rm up}=8$~M$_\odot$ end their life as core collapse or type-II\index{type-II supernovae} supernovae (SNe~II) (see~\cite{Smartt2009} and references therein). Under this assumption, the total amount of energy (typically $10^{51}$~erg per SN) that each star particle can release in the surrounding gas can be calculated. Within the approximation that the typical lifetime of massive stars which explode as SNII does not exceed the typical time step of the simulation, feedback energy and metals are injected by the dying star in the so-called ``instantaneous recycling\index{instantaneous recycling approximation} approximation'', i.e. in the same time step. 
On the other hand, type-Ia\index{Type-Ia supernovae} SNe (SNe~Ia) are believed to arise from the thermonuclear explosion of white dwarfs (see~\cite{Hillebrandt2000} and references therein). They lead to an explosion which is significantly delayed with respect to the time of creation of the star particle, as the white dwarf\index{white dwarf} in a binary system has to accrete matter from the companion and reach the mass threshold for the onset of thermonuclear burning.
Stars in the asymptotic\index{asymptotic giant branch (AGB)}\index{AGB, see asymptotic giant branch} giant branch (AGB) experience significant stellar mass loss and contribute significantly to the nucleosynthesis\index{nucleosynthesis} of heavy elements (see~\cite{Herwig2005} and references therein). 

Cosmological hydrodynamical simulations which feature an accurate model for stellar evolution and chemical enrichment can evaluate the number of stars aging and eventually exploding as SNe, as well as the amount of metals polluting the surrounding ISM. Besides assuming an IMF, commonly adopted models (e.g. \cite{tornatore07}) include stellar lifetimes, delay functions and table of stellar yields to calculate the chemo-energetic imprint mainly left by SNe~Ia, SNe~II, and AGB stars.

Chemical evolution models account for the evolutionary timescales of stars with different masses by adopting mass-dependent lifetime functions (i.e. functions that describe the age at which a star of mass~$m$ dies, e.g. \cite{PadovaniMatteucci1993, 1989A&A...210..155M, 1997ApJ...477..765C}). Stars with an initial mass larger than $M_{\rm up}$ and lower than $M_{\rm sup, SNII}=40$~$M_\odot$ are assumed to end their life exploding as core-collapse SNe, while stars that are more massive than $M_{\rm sup, SNII}$ are assumed to implode in BHs directly, and thus do not contribute to further chemical enrichment or stellar feedback energy.

A fraction (usually constant) of stars relative to the whole mass range is assumed to be located in binary systems that are progenitors of SNe~Ia. Assuming a stellar evolution scenario for SNe~Ia (e.g. according to the model by \cite{GreggioRenzini1983}) and adopting lifetime functions and delay time distribution functions, the rate of SN~Ia explosions can be computed. 

The production of different heavy metals by stars that evolve and eventually explode is followed by adopting stellar yields. They represent the ejected mass of different metal species $i$ produced by a star of mass $m$ and initial metallicity $Z$, i.e. $p_{Z_{i}}(m,Z)$. In general, predictions for at least three main processes are needed: the continuous mass loss of AGB stars, SNe~II and SNe~Ia. Tables of mass- and metallicity-dependent stellar yields are incorporated in the chemical evolution models of state-of-the-art cosmological hydrodynamical simulations. Predictions for stellar yield are still affected by uncertainties, mainly due to the poorly understood process of mass loss through stellar winds in stellar evolution models. Commonly adopted sets of stellar yields include: \cite{Thielemann2003} for SNe~Ia, mass- and metallicity-dependent yields by \cite{Karakas2010, Doherty2014a, Doherty2014b} for AGB stars, mass- and metallicity-dependent yields by \cite{woosley_weaver95, Nomoto2013, Romano2010, Romano2017} for SNe~II.
We refer the reader to (e.g. \cite{Greggio2005, Matteucci2003}) for a comprehensive review of the analytic formalism.

As a final caveat, we note that comparisons of results from simulations to observations of single stars are usually performed under the assumption that e.g., the metal content of a star particle in simulations statistically reproduces the mean metallicity of the SSP that the star particle samples. More accurate techniques are introduced in \cite{Valentini2018, Grand2018}.

\section{Stellar feedback}\label{sfr}

Stellar feedback is a key component of cosmological simulations of structure formation. Among its many crucial roles, it prevents overcooling at high redshift, distributes energy in the ISM, drives metals out of the star formation sites, and triggers galactic outflows, which guarantee a continuous interaction between the forming galaxy and its surrounding CGM. 
Since the understanding of the pc-scale physics responsible for launching and driving winds is still partial and however far from being implemented directly in cosmological simulations, these simulations have to resort to phenomenological prescriptions to capture the effects of galactic outflows. As for the sources of energy that power outflows and govern their kinematic, there are two commonly pursued approaches: winds can be either energy-driven or momentum-driven. For details about these two possible scenarios, we refer the reader to \cref{Ch_galform}, which provides complementary information on the specific topic of this sub-section. 
This outline is far from being complete: deeper insight can be gained e.g. from \cite{Mayer2008, Murante2015, Naab2017, Crain2023}.

A variety of stellar feedback schemes has been proposed in cosmological simulations: the numerical description of the energy injection from SN explosions in the ISM has indeed a strong impact on final results. How reliable different sub-resolution models are in capturing an effective description of feedback energy injection is still debated. 

SN energy can be distributed by star-forming gas elements to the surroundings in the form of thermal or kinetic energy. Injecting feedback energy as thermal, as in the pioneer model of \cite{Katz1996}, may have the limitation that the feedback usually does not result effective in driving energy far from regions where it is released. Gas elements surrounding those that provide energy have a typical temperature of $\sim 10^5$~K and density high enough to radiate energy almost immediately, their cooling time being short. As a result, feedback is not effective in counterbalancing cooling and in preventing excessive star formation. To overcome this weakness, a stochastic thermal feedback model has been proposed \cite{DallaVecchia2012}, where gas receiving thermal feedback energy is heated up to a temperature high enough to guarantee the effectiveness of feedback. Specifically, the aforementioned model assumes that thermal energy released by SN explosions is injected in the surrounding medium using a selection criterion: the gas particles that experience feedback have to be heated up to a threshold temperature, which is a parameter of the model (whose value is usually close to $10^7$~K). Such a temperature increase ensures that the cooling time of a heated particle is longer than its sound-crossing time, so that heated gas can effectively leave the star formation site before it radiates all the energy away. It is worth noting that stellar feedback is actually implemented as a thermal channel in this model: however, the effective outcome is a galactic wind, because thermal energy is converted into momentum and hence outflows originate. This model for stellar feedback is for instance adopted by the EAGLE and FLAMINGO simulations \cite{Schaye2015, Schaye2023}.

Otherwise, stellar feedback energy is provided in the form of kinetic energy and used to boost the velocity of surrounding gas elements (e.g. \cite{Springel2003, DallaVecchia2008}), which are kicked from their original position. They can eventually thermalise and radiate energy away, but later than in the thermal scenario by construction. To enforce the effectiveness of kinetic feedback schemes, particles that receive energy and sample galactic outflows (usually referred to as {\it{wind particels}}) are often decoupled from hydrodynamic interactions: in this way, they are prevented from being halted and from thermalising energy soon after they have been provided with.

An alternative stellar feedback model is represented by the so-called blast-wave feedback \cite{Gerritsen1997, Stinson2006}. Within this scheme, eligible particles are provided with thermal feedback energy, but are then prevented from cooling for a short period of time (typically few tens of Myr). The physical motivation 
behind this prescription stems from the evolution of a SN remnant in the ISM \cite{Clarke2007}. As soon as an exploding SN drives a blast wave, this undergoes a first phase of free expansion, followed by an adiabatic stage (the Sedov-Taylor phase) where radiative losses are negligible, before entering the radiative phase. Temporarily disabling cooling mimics the unresolved adiabatic phase, whose duration is estimated to be of order $30$~Myr; afterwards, gas is allowed to cool again. Alternatively, the switch off of cooling can be explained by assuming that the energy released by SN explosions generates turbulence at unresolved scales and is partially dissipated over few tens of Myr, thus hindering gas cooling \cite{Thacker2000, Governato2007}. When implemented in cosmological simulations, this feedback prescription results effective in avoiding excessive star formation and also succeeds at producing a two-phase ISM, whose gas components are not in pressure equilibrium locally.

Feedback processes through which massive stars can affect the reservoir of gas of a galaxy stem not only from the energy deposition and momentum injection following SN~explosions, but also from the ionizing effect that massive stars have before exploding (usually referred to as early stellar feedback; e.g.~\cite{Stinson2013}) and from stellar winds (e.g.~\cite{Fierlinger2016, Fichtner2022}). 
Specifically, the early stellar feedback by \cite{Stinson2013} represents a UV ionization source, that increases the surrounding gas temperature and supplies heating and pressure support. This feedback channel pre-processes the star-forming ISM and facilitates the effectiveness of SNe at regulating star formation. In their simulations, \cite{Stinson2013} show how early stellar feedback helps in suppressing SF at high z (see also the NIHAO simulations to appreciate how early stellar feedback operates \cite{Wang2015, Obreja2019}).

In addition, the energy released by SNe~II is not expected to be constant across cosmic time nor in all the star-forming regions, and theoretical models predict that SNe exploding in almost pristine or weakly enriched environments provide the ISM with a larger amount of stellar feedback energy.  
The idea of differentiating the outcome of stellar feedback according to the physical properties of the star-forming ISM has been already pursued in cosmological simulations, and a few effective prescriptions have been proposed. The basic idea consists in adopting a non-constant value of the stellar feedback efficiency (i.e. the fraction of energy provided by each SN that is actually coupled to the surrounding gas as feedback energy). 

A metallicity- and density-dependent stellar feedback efficiency has been introduced in the EAGLE simulations \cite{Schaye2015, Crain2015}. There, the aforementioned efficiency decreases with gas metallicity while increasing with gas density. The adoption of a similar parametrization has a twofold reason: $i)$ radiative losses are expected to increase with increasing metallicity; $ii)$ energy losses in high-density, star-forming regions can make the stellar feedback too inefficient, and have to be counterbalanced. 
Being the gas metallicity lower at higher redshifts, when also higher densities are usually reached in the star-forming ISM, such a prescription yields a redshift-dependent stellar feeback efficiency. The latter ranges between asymptotic values, which are parameters of the model tuned to reproduce low-redshift observables, e.g. the galaxy stellar mass function. 

\begin{figure}
 \centerline{\includegraphics[scale=0.4]{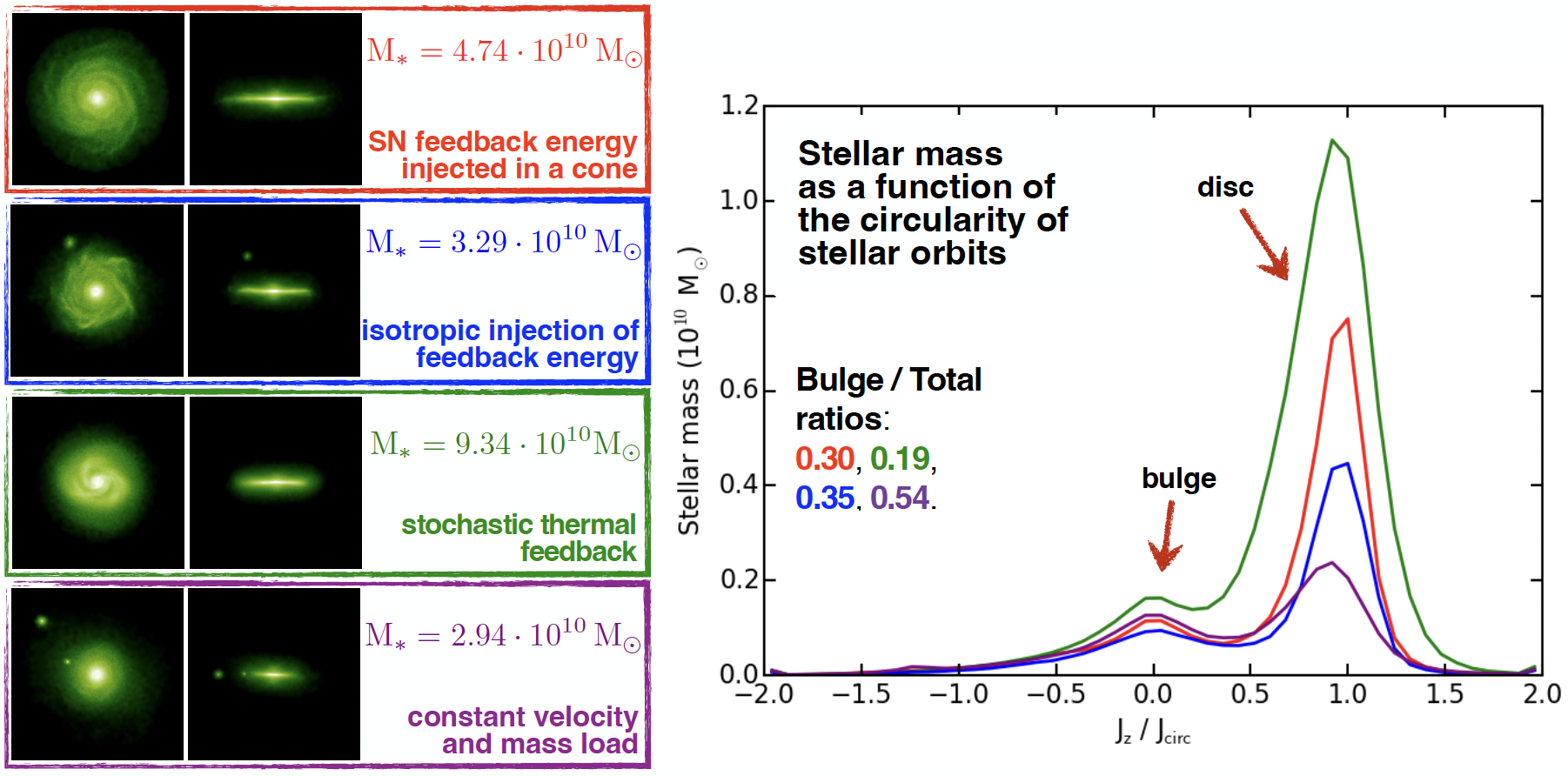}}
 \caption{Sensitivity of the final properties of simulated galaxies to the numerical prescriptions adopted to generate stellar feedback. Figures on the left show projected stellar density maps (face-on and edge-on) at redshift $z=0$. Four models of galactic outflows are implemented within the same code (Gadget3) and used to simulate the formation of a Milky Way-like galaxy: stellar feedback energy injected either within a cone (red) or isotropically (blue), a stochastic thermal feedback scheme (green), and a model where galactic outflows have constant velocity and mass loading factor (purple). The galaxy stellar masses listed for each model are computed within one tenth of the galaxy virial radius. On the right, the same finding is quantified by analysing the stellar mass in the disc and in the bulge components of the simulated galaxies: a kinematic decomposition based on the circularity of the stellar orbits -- through the ratio of specific angular momenta considered in the x-axis -- is employed. Simulations from \cite{Valentini2017}.\label{figs:stellar_feedback_models}}
\end{figure}

A similar parametrization is adopted in the Illustris-TNG simulation \cite{Pillepich2018}, too. 
They assume that the wind energy available to a star-forming resolution element is metallicity-dependent.  SNe II release indeed a feedback energy which spans the range $(0.9 - 3.6) \times 10^{51}$~erg, according to the metallicity of the star-forming gas cell in which they are expected to explode -- the lower the metallicity, the larger the energy budget.  \cite{Pillepich2018} demonstrate how the metallicity-dependent stellar feedback energy modulation has an impact both on the stellar-to-halo mass relation and on the galaxy stellar mass function, final results strongly depending on the tuning of the parameters.  They show that the stellar mass of a $\sim 10^{12}$~M$_{ \odot}$ halo at $z=0$ can vary by up to a factor of $\sim 2$.

At a higher level of complexity, though in simulations targeting smaller volumes, \cite{Maio2010} first modelled the simultaneous evolution of different stellar populations in cosmological simulations. Their runs feature stellar feedback dependent on the underlying gas metallicity and stellar population, where the injected energy and the assumed stellar yields depend on Population III versus Population II regimes, i.e. pair-instability SNe versus SNe.  Interestingly, \cite{Maio2016} included the effect of self-consistently coupled radiative transfer, concluding that radiative feedback from massive Population III stars could be a viable justification for powerful feedback in pristine environments. 

Motivated by the aforementioned findings and by theoretical studies (e.g., \cite{Bromm2004, Ciardi2005} for reviews) that suggest that hypernovae and SNe~II of Population III stars can release energy higher than in the present-day Universe by more than a factor of $\sim10$, \cite{Valentini2023} introduced an effective low-metallicity feedback. In this implementation of metallicity-dependent stellar feedback, they accounted for the effect of the explosion of hypernovae and SNe II in weakly-enriched environments, by boosting the energy released by SNe exploding in an almost pristine ambient medium (i.e. in an ISM with average metallicity below a threshold).

The impact of different sub-grid prescriptions accounting for the same physical process (stellar feedback and SN-triggered galactic outflows in this specific case) on final results is striking, though often overlooked.
\cite{Valentini2017} investigate the impact of stellar feedback modelling on the formation and evolution of a disc galaxy, by performing a suite of cosmological zoom-in simulations of a Milky Way-size halo. They show how sensitive the general properties of the simulated galaxy are to the way in which stellar feedback triggered outflows are implemented, comparing results obtained by adopting different state-of-the-art stellar feedback models (see \fref{figs:stellar_feedback_models}). 
We refer the reader to \cite{Scannapieco2012, Agora2021} for more extended comparison campaigns, i.e. the Aquila and the Agora comparison projects (see also \fref{figs:nifty}).
Interestingly, \cite{Crain2023} also show how making different assumptions as for the stellar feedback efficiencies in the EAGLE and Illustris-TNG simulations reflects on the predicted galaxy stellar mass functions (see also \cite{Crain2015, Pillepich2018}).

\section{Supermassive black holes in cosmological simulations}\label{Ch3_BHs}

Observations suggest that almost every galaxy hosts a supermassive ($10^8 \div 10^{10}$~M$_{\odot}$) black hole\index{black hole} (SMBH) in its innermost regions (e.g., \cite{Magorrian1998, Ferrarese2000, Gebhardt2000, Merritt2001, Tremaine2002, Marconi2003, Haring2004, KormendyHo2013, Gaspari2019}). A fraction of these BHs exhibits ongoing activity and is called AGN (active galactic nuclei). Evidence of past and ongoing activity is observed for instance in the X-ray images of elliptical galaxies and galaxy groups and clusters, where AGN imprints often appear as depressions and ripples. Among many, a well-known example is represented by the composite (X-ray, radio, and visual) image of the MS~0735+7421 galaxy cluster (e.g. \cite{Gitti2012}): giant X-ray cavities are filled with radio emission, and surrounded by a cocoon shock clearly visible in the Chandra image as an elliptical edge.

Before discussing AGN feedback (see Section~\ref{Ch3_agn_feedback}), we are now recalling the main features of BHs in cosmological simulations, focusing on BH seeding and repositioning, BH-BH mergers, and AGN feeding.

The majority of cosmological simulations of galaxy and galaxy cluster\index{galaxy cluster} formation that also include BHs and the associated AGN feedback are based on the seminal BH model by \cite{Springel_BHs, dimatteo05}, or follow its spirit. Therefore, this model offers the base to discuss a few key aspects of the treatment of BHs in cosmological simulations, and to highlight recent improvements.

In cosmological simulations, BHs are usually described as sink\index{sink particle} particles: these are collisionless particles that can absorb neighbour elements (originally implemented to allow the removal of surrounding particles in dense regions, e.g. \cite{Bate1995}) and that have fundamental properties like the accretion rate\index{accretion rate} which can be linked directly to observables. 

BH particles are introduced in massive haloes at relatively high-redshift in cosmological simulations, and they are then allowed to grow and increase their initial or {\sl {seed}} mass.
The aforementioned, commonly quoted mass $M_{\bullet}$ is the theoretical mass of the BH, modelled at the sub-resolution level, as opposite to its dynamical mass, i.e. the actual gravitational mass of the BH particle.
As we are still lacking a solid understanding of the formation of first SMBHs (see e.g., \cite{Bromm2003, Begelman2006, Mayer2010, Volonteri2012, Maio2018} for possible pathways) and the resolution needed to take the physics of any seed formation scenario into account, BHs are first inserted according to seeding prescriptions (see also \cite{DiMatteo2023}). 

Seeding prescriptions usually assume that new BHs (massive seeds of $\sim 10^4 \div 10^6$~M$_{\odot}$) are introduced in haloes which meet some criteria and do not have already BHs. 
New BHs are seeded if: $(i)$ the halo mass -- commonly estimated by means of a FOF (Friend-Of-Friend, \cite{Davis1985}) algorithm -- is larger than a threshold (e.g.,  \cite{Vogelsberger2014, Schaye2015, Pillepich2018, Schaye2023}); $(ii)$ the stellar mass exceeds a given value (e.g., \cite{Dave2019}); $(iii)$ the stellar mass and gas to stellar mass fraction are larger than given thresholds; $(iv)$ based on gas properties, i.e. gas density, velocity dispersion and/or metallicity (e.g., \cite{Taylor_Kobayashi2014, Habouzit2017, Wang2019, Dubois2021}).
BH mass at seeding can be either constant (e.g., \cite{Vogelsberger2014, Schaye2015, Pillepich2018, Schaye2023, Dave2019}) or scaled according to e.g., the M$_{\rm BH}$/$\sigma$ or M$_{\rm BH}$/M$_{\ast}$ scaling relations (e.g., \cite{Dolag2015}). 
Additional details involve the type of particles on which the FOF algorithm is performed, whcih can be for instance DM particles only (e.g., \cite{DiMatteo2008}) or stellar particles only (e.g., \cite{Dolag2015}). 
Besides, BH can be seeded at the position of the densest gas particle in the halo (e.g., \cite{DiMatteo2008}), of the star particle with the largest binding energy (e.g., \cite{Dolag2015}), or of the star particle closest to the centre of mass of the structure (e.g., \cite{Dave2019}).

In cosmological simulations, each BH undergoes a specific evolution as an individual particle, at variance with the coarse-grained representation of collisionless fluids through particles in N-body simulations. As a result, the dynamics of BHs fails to be accurately captured and numerical artefacts can affect BH motion (see e.g., \cite{Wurster2013}). Spurious displacements of BHs, which often occur due to scattering between particles in high-density environments and numerical heating, have dramatic consequences, such as e.g., artificial presence of wandering BHs, incorrect description of BH-BH mergers, BHs producing feedback off-centre with respect to their galaxy hosts.
To avoid these artefacts, BH dynamics is commonly controlled through different approaches: re-positioning via pinning on e.g. minimum potential (e.g., \cite{Vogelsberger2014, Schaye2015, Pillepich2018, Dave2019}), boosted dynamical mass (e.g., \cite{Bassini2020}), boosted dynamical mass and dynamical friction (e.g., \cite{Dolag2015, Dolag2023}). We refer the reader to \cite{DiMatteo2008, Debuhr2011, Hirschmann, Tremmel2015, ragone18, Chen2022, DiMatteo2023, Damiano2024, Genina2024} for details and recent improvements.

BHs grow because of gas accretion and mergers with other BHs.
As for the latter channel, BHs are expected to merge when their host galaxies and their haloes merge to form a single structure. 
The commonly pursued approach \cite{dimatteo05, Springel_BHs} in cosmological simulations assumes that two BH particles merge quickly if their distance approaches the spatial resolution of the simulation (or a small multiple of it). The force resolution set by the gravitational softening determines indeed the minimum scale above which gravitational interactions can be properly followed. Additional conditions involving e.g. the merging BH relative speed have been implemented \cite{DiMatteo2008}. 
The two BHs are eventually merged into a single BH particle, with their masses combined.

As for AGN feeding, gas accretion onto a BH of mass $M_{\bullet}$ is calculated according to the Bondi\index{Bondi} formula \cite{Hoyle,Bondi,Bondi_Hoyle}, multiplied by a so-called boost factor $\alpha$: 
\begin{equation}
\dot{M}_{\mathrm{B}} = \frac{4 \pi \, \alpha \, G^2\, M_{\bullet}^2 \, \langle \rho \rangle}{(\langle
c_\mathrm{s}\rangle^2 +\langle v\rangle ^2)^{3/2}} \,\, . 
\label{accretion_rate_mean}
\end{equation}
Here, $G$ is the gravitational constant,  $\langle\rho\rangle$, $\langle v\rangle$, and $\langle c_s\rangle$ are mean values at the scale resolved by the hydrodynamical simulation: for example, they are computed using kernel weighted estimates in the case of SPH. 
The BH accretion rate $\dot{M}_{\mathrm{B}}$ is commonly capped to the Eddington acccretion rate $\dot{M}_{\rm Edd}$\footnote{  $\dot{M}_{\rm Edd}=(4\pi\:G\;m_p\; M_{\bullet})/(\sigma_T\;c\;\epsilon_r)$, with $m_p$ the proton mass, $\sigma_T$ the Thompson cross section, $c$ the speed of light and
$\epsilon_r$ the radiative efficiency, typically assumed to be $\approx0.1$.} (but see \cite{Regan2019, Dave2019} for examples of $\dot{M}_{\bullet}$ which breaches the Eddington limit).

The boost factor $\alpha$ has been originally introduced \cite{Springel_BHs} to account for the limited resolution in simulations, which leads to smaller densities and larger temperatures near the BH (and thus to an underestimate of $\dot{M}_{\mathrm{B}}$). A typical value is $\alpha =100$. 
Several studies adapt the BH model by using a boost factor which depends on resolution \cite{Choi,Choi_2013}, density \cite{Booth}, or pressure \cite{Vogelsberger2013}. Other simulations instead limit the BH accretion rate by taking into account the angular\index{angular momentum} momentum of accreting gas \cite{Rosas_Guevara, Valentini2020}. 
High-resolution simulations of BH accretion on sub-kpc scales \cite{Gaspari} found that a boost factor of order of 100 is suitable when including cooling and turbulence\index{turbulence}, while pure adiabatic accretion suggests boost factors smaller by an  order of magnitude. Hence, advanced models distinguish between hot and cold gas accretion and use different boost factors for the two components \cite{Steinborn2015}, or can even get rid of fudge factors (e.g., \cite{Pillepich2018, Valentini2020}).

\begin{figure}
 \includegraphics[width=1.0\textwidth]{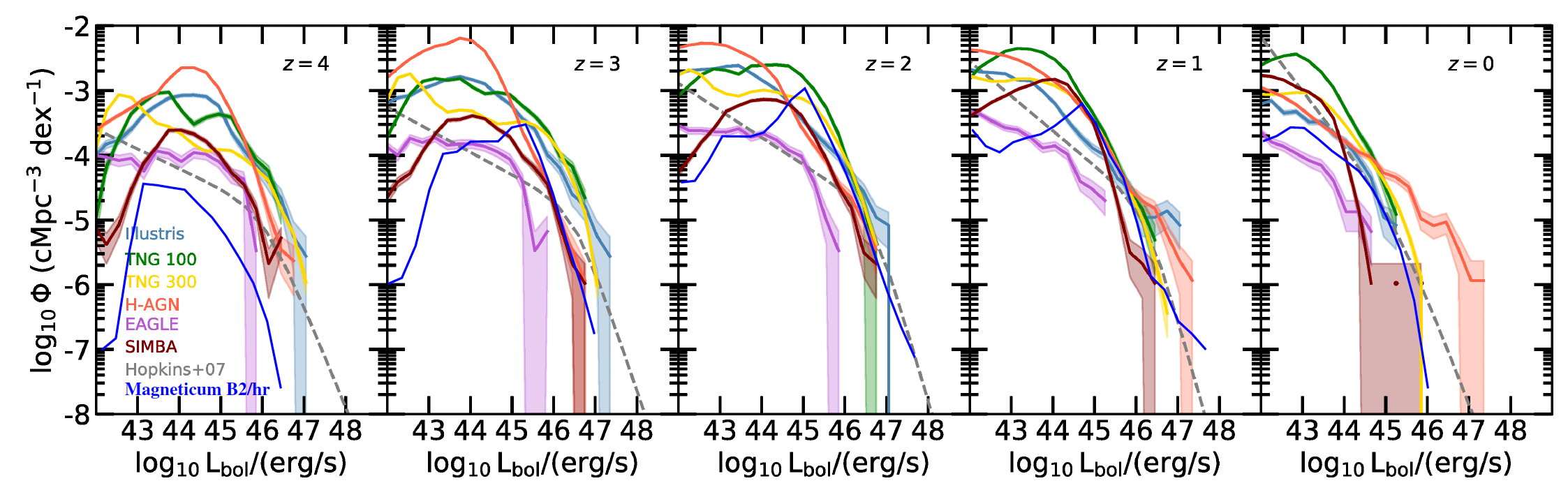}
 \caption{Comparison of the evolution of the observed AGN luminosity function with the predicted one from different simulations incorporating various AGN models (as described in Section~\ref{Ch3_agn_feedback}). Predictions from simulations are very sensitive to the details of the sub-grid models. Adapted from \cite{Habouzit2022}, and extended by including results from Magneticum, as shown in \cite{Hirschmann, Biffi2018agn}.
\label{fig:agn_lum}}
\end{figure}

By exploting the BH accretion rate $\dot{M}_{\bullet}$, the bolometric luminosity $L_{\mathrm bol}$ can be associated to each BH in the simulation. A common assumption consists in following \cite{Churazov2005} to distinguish between high and low BH accretion state in terms of the Eddington ratio $f_{\rm Edd} = \dot{M}_{\bullet}/ \dot{M}_{\rm Edd}$ (i.e. the ratio between the BH and the Eddington accretion rates) and compute:
\begin{equation}
L_{\mathrm bol} = 
\left\{ 
\begin{array}{ll} 
 10 \, ( \epsilon_r \; c)^2 \, \dot{M}_{\bullet}  & f_{\mathrm Edd}>0.1  \\ 
 (10 \, \epsilon_r\; c)^2 \, f_{\mathrm Edd}\; \dot{M}_{\bullet}\;\; & f_{\mathrm Edd}\leq0.1  
\end{array} \right. 
\end{equation}
(see \cite{Biffi2018agn, Habouzit2022}). 
In contrast to the original model \cite{dimatteo05, Springel_BHs}, modern implementations often correct the accretion rate $\dot{M}_{\bullet}$ of the BH by a factor $(1-\epsilon_r)$. As a consequence, $\,L_{\mathrm bol} = \epsilon_r / (1-\epsilon_r) \, \dot{M}_{\bullet} \,c^2 \,$, and the energy  radiated away during the accretion process is taken into account. 
Different prescriptions for the accretion rate, in combination with the change of the ISM/IGM properties produced by the combined action of all the sub-grid processes, typically leads to significant variations in the predicted evolution of the AGN luminosity function among the various state-of-the-art simulations. This is summarized in \fref{fig:agn_lum}, where we can appreciate how predictions from simulations can strongly deviate from the observed AGN luminosity function across cosmic time (see also discussion in \cite{Habouzit2022}).

\section{AGN \index{AGN feedback} feedback in state-of-the-art cosmological simulations}\label{Ch3_agn_feedback}

Observations allow us to appreciate how AGN feedback develops with different mechanisms (for instance inflating bubbles or launching outflows), with different appearances, in a variety of systems.
Besides, it is a recurrent process, with each AGN burst leaving a clear signature in the system.
Additional evidence which adds complexity to the comprehension and modelling of AGN feedback is the presence of multiphase gas in and around galaxies.
Multiwavelength observations reveal the presence of gas spanning a wide range of densities, temperatures, and ionisation states in galaxies. Multiphase gas is present not only in spiral galaxies, which are systems rich in cold gas, but also in ellipticals and in the innermost regions of galaxy groups and clusters, which are environments known to be dominated by X-ray emitting, hot gas.

Observations (\cite{Russell2013, Mezcua2014} among other works) also provide evidence for different regimes of BH accretion and feedback. In a simplified context, the AGN activity operates through (at least) two distinct phases: radiative or quasar mode, and kinetic or radio mode (\cite{Fabian2012} for a review). The radio-mode is characterized by large radio jets generating hot X-ray cavities, whereas in the quasar-mode the emission is dominated by the accretion disc, and feedback energy is mainly radiated away. This distinction has been theoretically modelled \cite{Churazov2005} by describing AGN feedback with two components: radiation and mechanical outflows (\cite{Merloni_Heinz2008} considered three different regimes for BH accretion in their model). In these models, the amount of energy associated with each component depends on the Eddington ratio $f_{\rm Edd}$ (see Section~\ref{Ch3_BHs}). 
Not only are simple theoretical models like the two aforementioned ones supported by observations, but they are also corroborated by other independent analytical models (\cite{Yuan_Narayan2014} and references therein), which showed how it is possible to associate different BH accretion rates to changes in the radiative efficiency and to transitions to different types of accretion discs. 
In addition, observations \cite{Davis, Chelouche} suggest that the radiative efficiency does not only correlates with the BH accretion rate, but also with the BH mass (see also \cite{Steinborn2015}).

However, a two-mode AGN feedback is only an approximation to the smooth transition which is observed (e.g., \cite{Russell2013, Heckman2014}) and also theoretically expected (e.g., \cite{Fender2004, Tchekhovskoy2012, Yuan_Narayan2014, Sadowski2016}).
Reality is in fact more complex than the aforementioned simplified frameworks, and AGN feedback operates through a variety of modes, which often occur simultaneously. Among the main AGN feedback mechanisms, there is the preventive (or preventative) mode, where feedback prevents the gas from being accreted or from effectively cooling; the ejective mode, in which gas is removed from the innermost regions of forming structures where SF occurs; AGN feedback can suppress the SF efficiency, mainly via turbulence and ISM heating; it can operate via radiation pressure on dust, which clears out gas from the BH vicinity; or AGN can act in the maintenance mode, by keeping quiescent an already quenched system. 
Besides all the aforementioned modes in which BH feedback is negative, i.e. it produces an overall suppression of the SFR, AGN feedback can also be positive and locally enhance the SF efficiency for a given period of time, mainly via ISM overpressurization. 
Different regimes have to be accounted for to address the overall AGN feedback problem, always considering that 
AGN and their host galaxies are embedded in a DM halo.

Another key characteristic of AGN is that it develops across a large dynamical range of scales, and that it impacts on the evolution of gas with a range of densities and temperatures. This can occur from nuclear scales to galaxy cluster scales, moving through the ISM, the CGM, the intra-group medium and the ICM. 
This large range in spatial scales corresponds to an equivalently wide range in temporal scales (e.g., \cite{Harrison2024}).

State-of-the-art cosmological simulations are still far from fully capturing the complexity. Starting from the seminal works by \cite{dimatteo05, Springel_BHs}, a number of prescriptions for modelling AGN feedback have been conceived to account for SMBH feedback in cosmological simulations. Following the two aforementioned papers, many simulations assume that the AGN feedback\index{feedback} energy that a SMBH accreting at a rate $\dot{M}_{\bullet}$ releases per unit time can be cast as:
\begin{equation}
\dot{E} = \epsilon_\mathrm{f} \epsilon_\mathrm{r} \dot{M_\bullet} c^2,
\label{feedback_energy_old}
\end{equation}
where $\epsilon_\mathrm{r}$ and $\epsilon_\mathrm{f}$ are the radiative\index{radiative efficiency} and the feedback efficiencies, respectively. 

A constant value for the radiative efficiency $\epsilon_\mathrm{r}$ is often used. It is often set to $0.1$ \cite{Springel_BHs, Dubois2016}, which corresponds to the mean value for the radiatively efficient accretion onto a non-spinning, Schwarzschild BH \cite{Shakura1973, Novikov1973, Yu_Tremaine2002}. Recent works (e.g., \cite{Dubois2021, Sala2024}) included a BH spin-dependent $\epsilon_\mathrm{r}$.

The feedback efficiency $\epsilon_{\rm f}$ quantifies the fraction of energy radiated from the BH which is actually coupled to the\index{ISM}. Values commonly assumed in cosmological simulations span the range~$\sim 10^{-4} - 0.2$ (e.g., \cite{Ostriker2010, Dubois2012, Costa2014MNRAS439, Hirschmann, Lupi2019, Pillepich2018, Trebitsch2020}). This efficiency is often tuned in order to match the normalisation of the BH to stellar mass relation \cite{Magorrian1998, KormendyHo2013}. 
The value of $\epsilon_{\rm f}$ can be either constant (e.g., \cite{Springel_BHs, Schaye2015}, or dependent on the AGN mode. Motivated by the observational evidence of more powerful outflows associated with SMBHs accreting at lower rate \cite{Churazov2005, Merloni_Heinz2008, Best_Heckman2012}, \cite{Sijacki} introduced a steep transition of the feedback\index{feedback efficiency} efficiency between quasar-mode\index{quasar-mode} and radio-mode,\index{radio-mode} (see also \cite{Fabjan2010, Dubois2012_BH, Hirschmann, Vogelsberger2013, Dubois2016, Dave2019}). Even within the same (radio) mode, $\epsilon_{\rm f}$ can change according to the ISM density \cite{Weinberger2017, Pillepich2018}.

AGN feedback energy can be deposited by means of different numerical prescriptions: either purely in the form of thermal or kinetic energy, or it can be used to inject bubbles, or according to different channels depending on e.g., the BH accretion rate.
Among the several possible AGN feedback models, it is worth highlighting a number of schemes that modern cosmological simulations adopt.

\begin{figure}
 \centerline{\includegraphics[trim=10 710 10 1,clip,width=1.25\textwidth]{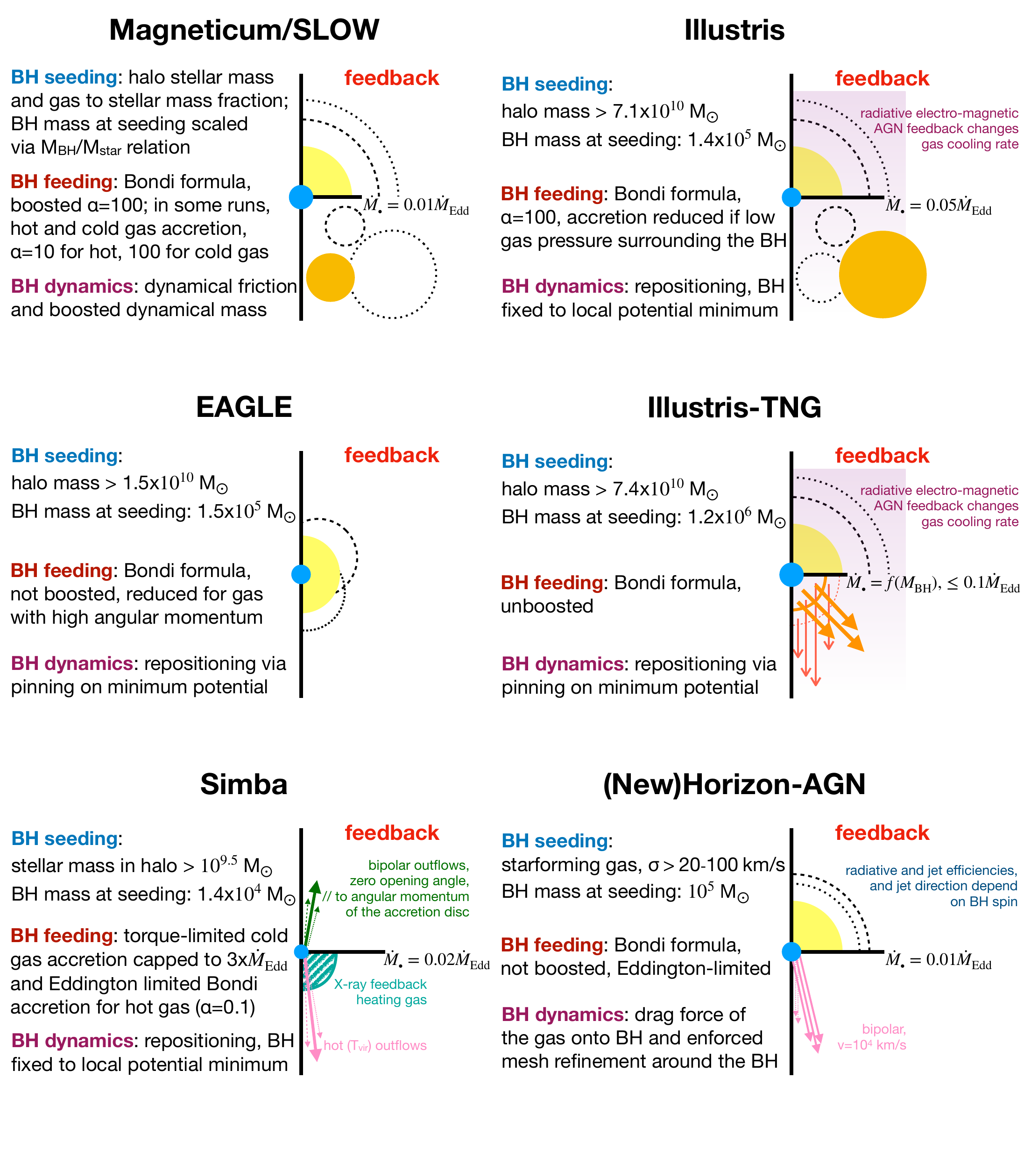}}
 \caption{The six cartoons summarise the main properties of BH seeding and accretion, and sketch the implementation of AGN feedback in six among the most common state-of-the-art cosmological simulations. In the right part of each panel, we highlight the BH accretion rate threshold above (top) or below (bottom) which different AGN feedback prescriptions are used, should a two-mode feedback be present in that simulation suite.\label{figs:agn_models_cartoon}}
\end{figure}

Following the original idea of \cite{dimatteo05, Springel_BHs}, the AGN feedback energy can be simply dumped thermally: in this way, the BH distributes the AGN feedback energy (eq.~\ref{feedback_energy_old}) to nearby resolution elements, possibly in a kernel-weighted fashion. The thermal energy received is used to increase the internal energy of particles or cells surrounding the BH, which increase their temperature as a consequence. While being overall effective, this simple prescription often turns out to be inefficient and does not succeed at reproducing the variety of observations currently available at different redshift. Nonetheless, this prescription is still widely used in cosmological simulations, that often adopt it to describe the quasar-mode feedback only, along with a complementary model for lower BH accretion rates.
As an example, \cite{Vogelsberger2013, Hirschmann, Dubois2016, Pillepich2018, Dubois2021, Dolag2023} rely on this model to describe the quasar-mode feedback in their simulations. 

Indeed, in the Magneticum and in the SLOW simulation suites, they assume a two-mode AGN feedback: for high BH accretion rates, i.e. $f_{\rm Edd} > 0.01$, SMBHs experience a quasar phase and release thermal energy in the surrounding. On the other hand, they model the radio-mode feedback ($f_{\rm Edd} < 0.01$) through energy deposition by hot bubbles (following \cite{sijacki07, Fabjan2010}). In these simulations, radiative and feedback efficiencies are free parameters. Specifically, $\epsilon_{\rm f}$ is increased by a factor of 4 during the radio-mode feedback with respect to the quasar phase.
Interestingly, in a subset of the Magneticum simulations \cite{Steinborn2015}, they explore a different scenario for a two-mode AGN feedback, where hot bubbles are replaced by outflows. However, due to the limited resolution, the latter mechanical channel is numerically implemented as thermal feedback.
 
While the Illustris simulation \cite{Vogelsberger2013, Vogelsberger2014} implements AGN feedback in a similar way to Magneticum (i.e. thermal feedback for $f_{\rm Edd} > 0.05$ and bubble-like feedback below), the Illustris-TNG simulation substantially improves the modelling of the mechanical component of the BH feedback \cite{Weinberger2017, Pillepich2018}. Indeed, besides adding a BH mass dependence to the BH accretion rate threshold which distinguishes between quasar and radio mode, they explicitly model kinetic winds in the radio mode, by adding momentum with random injection directions to selected gas cells (see \cite{Weinberger2017} for details). 
Interestingly, both Illustris and Illustris-TNG include a radiative (electro-magnetic) AGN feedback via a phenomenological model. This latter feedback channel modifies the net cooling rates of halo gas. Even though it operates whatever the $\dot{M}_{\bullet}$, it proved effective only for BH accretion rates close to Eddington.

The EAGLE simulations \cite{Schaye2015} adopt a stochastic thermal feedback scheme, where accumulated AGN feedback is injected only when enough to heat the surrounding ISM up to an sufficiently-high temperature ($T \sim 10^{8.5} \div 10^9$~K), similarly as for the case of stellar feedback (see Secton~\ref{sfr}). 
This choice is also the fiducial option in the reference runs of the FLAMINGO suite \cite{Schaye2023}, although they foresee the possibility to adopt instead a jet-mode, kinetic AGN feedback. In the latter case, should jets be on, they are launched according to the BH spin direction. AGN feedback in these two simulation suites \cite{Schaye2015, Schaye2023} is always one-mode.

Simba simulation \cite{Angles-Alcazar_2017, Dave2019, Scharre2024} includes a two-mode AGN feedback: a radiative and a jet mode. BHs in the radio mode ($f_{\rm Edd} > 0.02$) promote AGN winds, with velocity proportional to the BH mass, that do not alter the temperature of the outflowing gas. During the jet-mode (for low $f_{\rm Edd}$ and if $M_{\bullet} > 10^{7.5}$~M$_\odot$), the temperature of the gas involved in outflows is instead raised to the virial temperature of the halo. Both types of winds are produced via momentum input. When the jet-mode is active, there is an additional heating channel, namely the X-ray feedback: it releases energy into the BH surrounding gas, which is supposed to originate from X-rays of the accretion disk.

In the (New)Horizon-AGN \cite{Dubois2012_BH, Dubois2014_spin, Dubois2016, Dubois2021} AGN feedback features two modes, depending on the BH accretion rate. During the radio mode, the BH powers jets, continuously releasing mass, momentum, and total energy into the gas. On the other hand, only thermal energy in supplied to the gas cells within a sphere centred on the BH when it is in quasar mode. As for the jet mode, BH jets are bipolar, launched with a constant velocity, and they deposit mass, momentum and energy within a cylinder. In the New Horizon-AGN simulation, radiative and feedback efficiencies are BH spin-dependent (both in quasar and radio mode).

\fref{figs:agn_models_cartoon} summarises the numerical prescriptions that some modern cosmological simulations adopt to model BH growth, AGN feeding and feedback. The six sketches in the aforementioned figure aim at providing a simplified yet immediate understanding and comparison among the various models.

The compilation of prescriptions for AGN feedback listed above is far from being complete.
Other works which is worth mentioning include e.g., \cite{Angles-Alcazar_2017_Fire, Henden2018, Tremmel2019, Blank2019, Ni2022}.

A few remarks on AGN feedback modelling in simulations follow, about: $(i)$ model parameter calibration; $(ii)$ the simplification underlying many of the state-of-the art models; and $(iii)$ comparison among different models.

Subgrid physics and in particular free parameters of the models, are usually tuned to reproduce observables. This is especially true for AGN feedback efficiencies (i.e. $\epsilon_{\rm f}$ and $\epsilon_{\rm r}$). 
The majority of cosmological simulations calibrate free parameters (possibly within ranges suggested by theory or phenomenology or data) by attempting to reproduce a number of observables. The most popular examples are the BH to stellar mass relation \cite{Magorrian1998, KormendyHo2013} and the galaxy stellar mass function (GSMF, e.g. \cite{Baldry2012}), usually at redshift $z=0$. Other calibration targets include: the galaxy mass-size relation, baryon or gas mass fractions, metallicity-mass and luminosity-mass relations. We refer the reader to \cite{Crain2015} for a more extended discussion. 
Common strategies to model calibration consist in performing a fit to selected data sets (e.g. EAGLE), to exploit trends in comparison to some observations (e.g. Magneticum, Illustris), to profit from machine learning (e.g. FLAMINGO) or from comparison to previous/scaled-down versions of similar simulations.

\begin{figure}
 \includegraphics[width=1.0\textwidth]{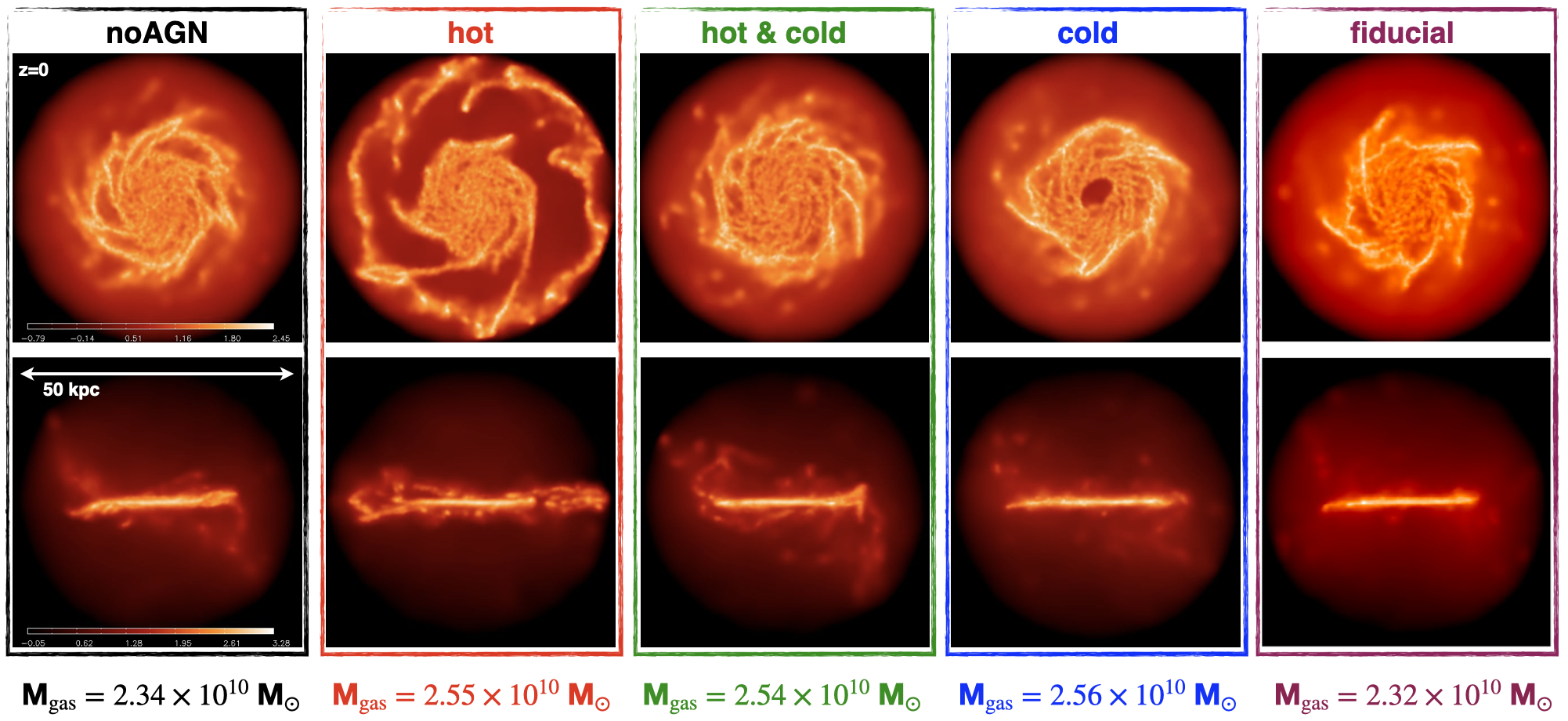}
 \caption{Impact of different ways of distributing AGN feedback energy on the final morphology of simulated disc galaxies. Besides the reference run where no BHs are included (first column), we show the final morphology of a suite of Milky-Way like galaxies which have been simulated assuming that AGN feedback energy $(i)$ is entirely distributed to the hot phase of the multiphase ISM (second column), $(ii)$ is evenly provided to hot and cold gas (third column), $(iii)$ is coupled to the cold phase only (fourth column), or $(iv)$ is supplied to hot and cold gas according to their physical properties (i.e., depending on the covering factor of the cold clouds, fifth column). Figure panels show projected gas density maps (top panels: face-on, bottom panels: edge-on) of five simulated galaxies, at redshift z = 0. The size of each box is 50 kpc a side. Colour bars encode the logarithm of projected densities (M$_\odot$/pc$^2$); panels in each row share the same colour bar. Shown are also the mass of gas within $0.1$ of the galaxy virial radius. Adapted from \cite{Valentini2020}. }
\label{fig:Muppi_AGN}
\end{figure}

There are as well different approaches when resolution changes. Weak convergence (as opposite to strong convergence, see \cite{Schaye2015}) is usually preferred, and for different numerical resolutions it is common practice to adapt at least a few values of the model parameters.

However, the choice of the observational data sets adopted for calibration does not impact parameter tuning only. Rather, it allows to understand the performance of different simulation predictions against various observations.
For instance, the Magneticum simulation succeeds at reproducing the physical properties of the ICM and of the intra-group medium better than other simulations (see \cite{Bahar2024} for an example), while simulations like Illustris-TNG, Simba and EAGLE usually outperform the competition as for predicting final properties of the galaxy population. This mainly stems from the calibration targets, i.e. ICM and stellar metallicity content, plus X-ray luminosity-mass relation for the Magneticum simulation versus GSMF and galaxy mass-size in the others.

\begin{figure}
 \includegraphics[width=1.0\textwidth]{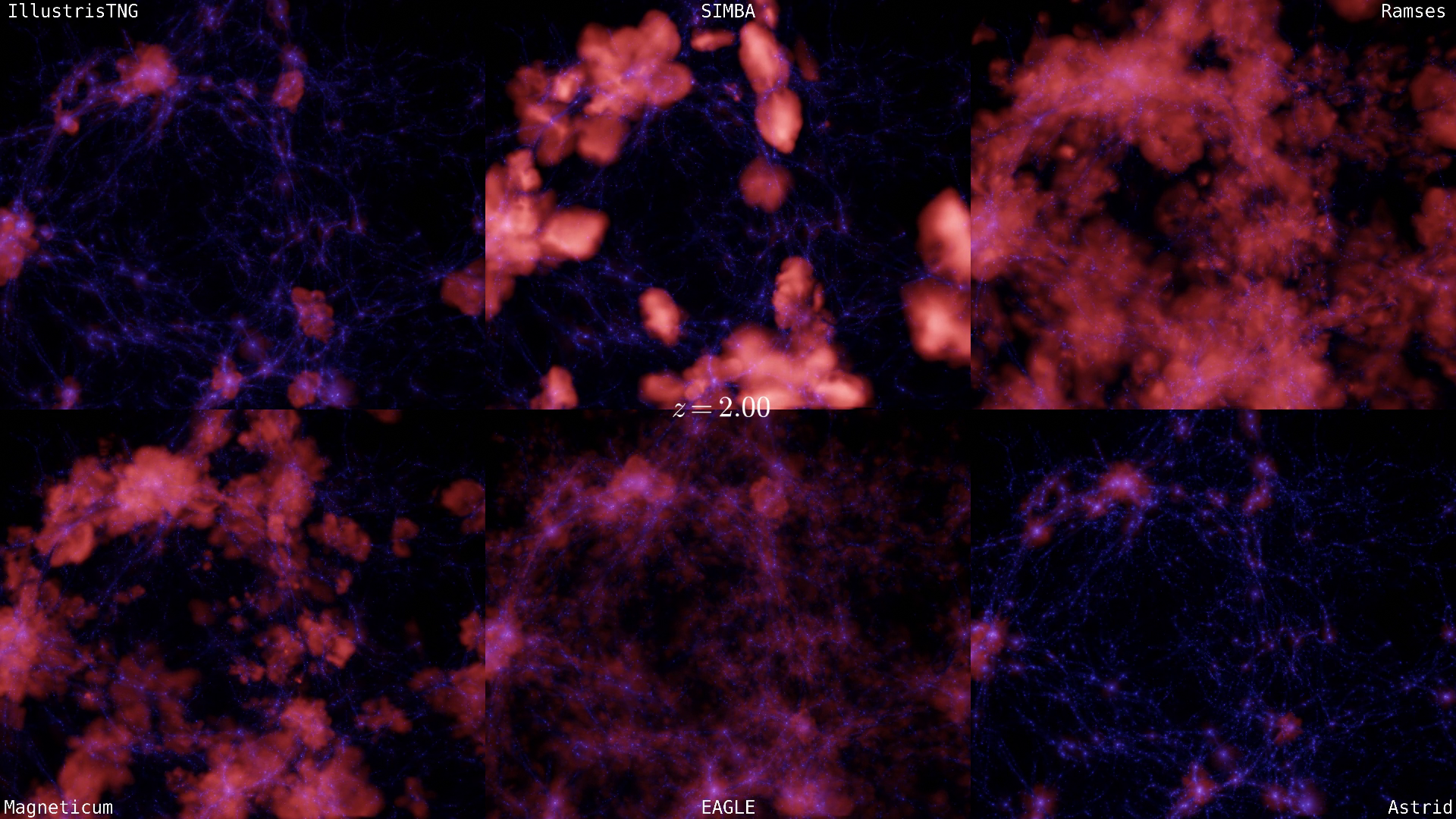}\\
 \includegraphics[width=1.0\textwidth]{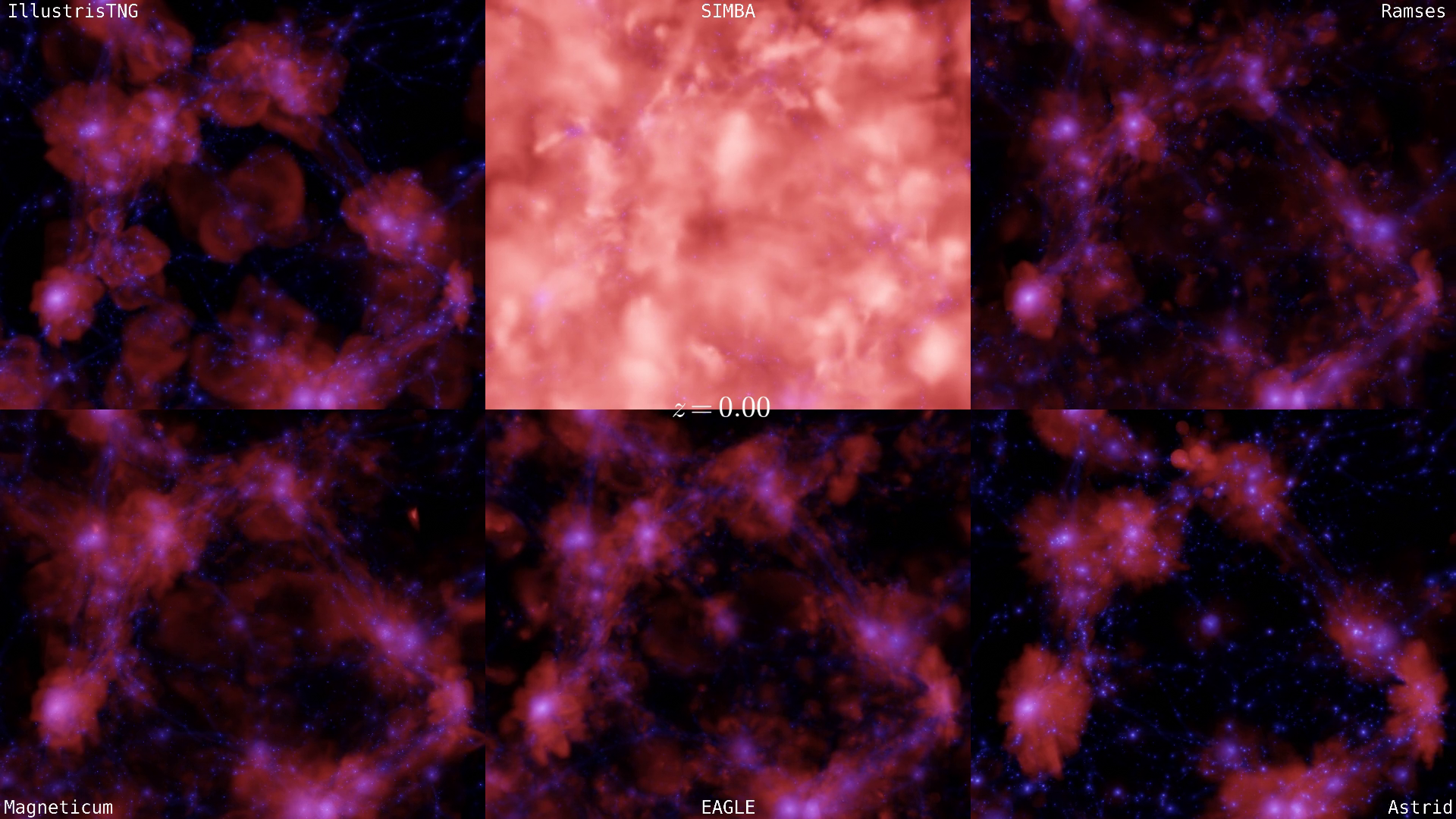}
 \caption{Comparison of the evolution of the ISM (the displayed density maps are color coded by temperature) within a small, cosmological volume evolved within the CAMELS project. Six different sub-grid models which have been used in recent simulation campaigns are compared (see labels). The upper 6 panels show the simulations at $z=2$, while the lower panels show the simulations at $z=0$, so as to also emphasize the different time evolution inherited in the different implementations. 
 Credits: F. Villaescusa-Navarro. 
}
\label{fig:camels}
\end{figure}

Among the many assumptions underlying any sub-grid model, one simplification consists in accepting that little is known about how AGN feedback energy couples with different phases of the ISM. 
\cite{Valentini2020} investigate how relevant it is to distribute AGN feedback energy to the individual phases of the ISM. Within the MUPPI sub-resolution model \cite{Murante2015, Valentini2017}, they show the impact of supplying feedback energy to hot or cold gas only, of evenly providing the two phases with BH feedback energy, and of coupling AGN feedback energy to the hot and cold gas according to their physical properties (i.e. the covering factor of cold clouds). They find that the numerical prescriptions adopted to couple AGN feedback energy to the ISM phases affect final BH and host galaxy properties (see \fref{fig:Muppi_AGN}). Interestingly, they assume that AGN feedback energy is used either to increase the hot gas temperature, or to evaporate molecular gas, instead of e.g. producing a deviation from the equilibrium solution of multiphase particles \cite{Springel_BHs}.

Finally, assessing the differences among different models for AGN feeding and feedback is not straightforward. We refer the interested reader to \cite{Habouzit2021} for a reference comparative study where the impact of various prescriptions for BH seeding, AGN feeding and feedback has been addressed. Different implementations are tested with the same cosmological simulation also in the CAMELS project \cite{Villaescusa-Navarro2021}. Here, \fref{fig:camels} (along with the associated movie) provides evidence of how various sub-grid models adopted in some state-of-the-art cosmological simulations predict differences as for the energy distribution and the evolution of the IGM/CGM.

\section{Current state and perspectives}

Even nowadays, the cosmological simulations used as theoretical counterparts for very large surveys (like EUCLID, DESI, LSST), for which extremely large volumes need to be sampled, are based on pure gravitational physics (e.g., \cite{2017ComAC...4....2P, 2021MNRAS.506.4210I}). These simulations are usually complemented by running semianalytic models\index{semianalytic models} (SAMs) of galaxy formation (e.g., \cite{Hirschmann2016}). While SAMs provide a realistic description of the properties of galaxy populations, they bring at best indirect information on the properties of the ISM/IGM\index{IGM}. In fact, they do not include a self-consistent treatment of gas dynamics and loosely capture the effects of baryons on structure formation, which is highly relevant for the study of environmental effects.

In the last decades, a growing number of different, large-scale, cosmological, hydro-dynamical simulations have been performed with varying resolutions and volumes covered (see initial \fref{fig:sims} in section \ref{ch3_intro}). One limitation of state-of-the-art cosmological simulations is that different sub-grid models often predict very similar results in direct observable properties, and the main differences among their outcome are often hidden in properties which are not directly or easily accessible (like the IGM/CGM of galaxies), and/or in the different evolution of them (see \fref{fig:camels}). This weakens the predictive power of individual simulations, and undermines the possibility to ascribe discrepancies among simulations to different numerical prescriptions adopted or to physical processes included. Comparing predictions from simulations with observations can help to constrain the theoretical modelling: however, evaluating differences fairly is not trivial. In fact, the comparison process often involves sophisticated techniques to create meaningfully mock observations (e.g., \cite{Biffi2012, Valentini2018}), and also has to cover multi-wavelength regimes and different components (e.g., stars, gas, dust, ...) simultaneously.

Several lines of refinement have been under development during the last years. They include, for instance, hyper-refinement in the innermost regions of simulated structures \cite{Curtis_Sijacki2015, Koudmani2019, Angles-Alcazar2021}, to increase the resolution in those regions which are crucial to better resolve to capture additional physics. 
Not only is higher resolution needed in the densest regions or around SMBHs, but it is fundamental also increase the spatial refinement within the virial radius, to capture CGM physics and evolution more accurately (e.g., \cite{van_de_Voort2019}). 
Also, a number of works have included a detailed modelling of sub-resolution accretion discs and BH spin modelling \cite{Dubois2014_spin, Fiacconi2018, Bustamante2019, Husko2022, Sala2024}, and have progressed in linking the accretion process onto SMBHs with the launch of AGN-triggered jets \cite{Bourne2017}. 
Furthermore, the modelling of SMBH outflows and jets has been enhanced \cite{Weinberger2017, Cielo2018}, and interests experiments involve the inclusion of spin-driven (Blandford-Znajek) AGN jets \cite{Talbot2022, Talbot2024}.
So far, these improvements have been validated mainly with dedicated and somehow ad-hoc setups, or included in cosmological simulations of smaller haloes or volumes. As the numerical advancement of the aforementioned improvements has been included in state-of-the-art codes, we expect that these new modules will be featured by upcoming large-volume simulations, and improve the accuracy of the numerical prediction in the near future.

While the predictive power of modern cosmological simulations is beyond discussion, still there are several caveats that cannot be overlooked. 
The majority of state-of-the-art cosmological simulations, for instance, assume that all haloes above a mass threshold host BHs, and adopt massive BH seeds to promote their early growth. Also, they commonly rely on Bondi accretion -- though with modifications -- to facilitate the initial growth of BHs and to make them easily reach the Eddington limit. 

In addition, the implementation of feedback processes is still far from reaching an adequate degree of complexity, and it is often driven by numerical effectiveness.
Besides, several feedback processes (e.g., early feedback, radiation from stars, stellar and AGN winds, radiation pressure on dust) which could hamper BH growth and reduce the SFR are often neglected. 
A number of relevant physics modules are still not included in the reference runs of the majority of cosmological simulations of large volumes, as their integration in the existing framework is not straightforward. As an example, it is worth mentioning: radiative transport, cosmic rays, magnetic fields, scenarios for DM alternative to the standard $\Lambda$CDM, chemical diffusion, just to name a few.
The need for higher spatial resolution makes it difficult to consistently account for all the relevant physics.

Finally, it is worth recalling that we have now entered the era of high-performance computing, with always growing computational power available from current and future HPC facilities. 
It is therefore of paramount importance to have numerical codes which are not only as complete as possible as for the inclusion of physical processes implemented, but also very efficient from the computational point of view. Codes are indeed supposed to be able to smoothly scale on state-of-the-art exascale infrastructures, and to efficiently exploit CPUs (Central Processing Units) and GPUs (Graphics Processing Units). 

The aforementioned efforts, together with the improvements in numerical methods, will allow us to study the formation process of cosmological structures with unprecedented detail, and to better link the small-scale physical processes of galaxy formation to the evolution of the large-scale structure and of the cosmic web.

\bibliographystyle{ws-rv-van}

\bibliography{NEW_03_chapter_second_edition}

\end{document}